\documentclass[a4paper,11pt]{article}
\usepackage[left=30mm,right=20mm,top=20mm,bottom=25mm]{geometry}
\usepackage{amsmath}
\usepackage{amsfonts}
\usepackage{amssymb}
\usepackage{amsbsy}
\usepackage{color,graphics}
\usepackage{setspace}
\usepackage{epsfig}
\usepackage{subfigure}
\usepackage{cite}
\usepackage{graphicx}
\usepackage{latexsym,multicol}
\usepackage{tcolorbox}

\graphicspath{{./LCE-instability-figs/}}
\begin{document}
	\title{\huge\textsf{\textbf{Instabilities in liquid crystal elastomers}}}
	\author{L. Angela Mihai$^{\dagger}$\qquad
		Alain Goriely$^{*}$\\ \\
		\textit{$^{\dagger}${School of Mathematics, Cardiff University,  Cardiff, UK}}\\
		\textit{$^{*}$Mathematical Institute, University of Oxford,  Oxford,  UK}\\ \\ \small{ORCID: L.A.M. (0000-0003-0863-3729); A.G. (0000-0002-6436-8483)}}
	\maketitle
	
	\begin{abstract}\vskip 6pt
		\hrule\vskip 12pt
	Stability is an important and fruitful avenue of research for liquid crystal elastomers. At constant temperature, upon stretching, the homogeneous state of a nematic body becomes unstable, and alternating shear stripes develop at very low stress. Moreover, these materials can experience classical mechanical effects, such as necking, void nucleation and cavitation, and inflation instability, which are inherited from their polymeric network. We investigate the following two problems: First, how do instabilities in nematic bodies change from those found in purely elastic solids? Second, how are these phenomena modified if the material constants fluctuate? To answer these questions, we present a systematic study of  instabilities occurring in nematic liquid crystal elastomers, and examine the contribution of the nematic component and of fluctuating model parameters that follow probability laws. This combined analysis may lead to more realistic estimations of subsequent mechanical damage in nematic solid materials.\\
	
		\noindent{\bf Impact statement:} Due to their complex material responses in the presence of external stimuli, liquid crystal elastomers have many potential applications in science, manufacturing, and medical research. The modeling of these materials requires a multiphysics approach, linking traditional continuum mechanics with liquid crystal theory, and has led to the discovery of intriguing mechanical effects. An important problem for both applications and our fundamental understanding of nematic elastomers is their instability under large strains, as this can be harnessed for actuation, sensing, or patterning. The goal is then to identify parameter values at which a bifurcation emerges, and how these values change with external stimuli, such as temperature or loads. However, constitutive parameters of real manufactured materials have an inherent variation that should also be taken into account. Hence, the need to quantify uncertainties in physical responses, which can be done by combining the classical field theories with stochastic methods that enable the propagation of uncertainties from input data to output quantities of interest. The present study demonstrates how to characterize instabilities found in nematic liquid crystal elastomers with probabilistic material parameters at the macroscopic scale, and paves the way for a systematic theoretical and experimental study of these fascinating materials.\\
		
         \vskip 6pt
		\hrule
	\end{abstract}
	

\section{Introduction}

Liquid crystal elastomers (LCEs) are advanced multifunctional materials that combine the flexibility of polymeric networks with the nematic structure of liquid crystals \cite{deGennes:1975,Finkelmann:1981:FKR}. Due to their complex molecular architecture, they are capable of exceptional responses, such as large spontaneous deformations and phase transitions, which are reversible and repeatable under certain external stimuli, namely: heat, light, solvents, electric or magnetic fields. These properties render them as promising candidates for future `animate materials', and could be harnessed for a range of technological applications including soft actuators and soft tissue engineering. Nevertheless, a better understanding of these materials is needed before they can be exploited on an industrial scale  \cite{deHaan:2014:etal,Jiang:2019:JXZ,Kuenstler:2019:KH,Mahimwalla:2012:etal,McCracken:2020:etal,Pang:2019:etal,Ube:2014:UI,Ula:2018:etal,Wan:2018.etal,Warner:2020,Wen:2020:etal,White:2018,White:2015:WB,Xia:2019:XHY}.

Since the early discovery of liquid crystalline solids, probing their intriguing material properties has been the focus of research laboratories around the world, and the importance of such essential work is hard to overstate. However, their accurate description can only be useful if fully integrated in a multiphysics framework combining elasticity and liquid crystal theories. Many nematic solids are synthesized as polydomains, where the liquid crystal mesogens are separated into different domains, and in every domain, they are aligned along a preferred direction,  known as the \textit{local director} \cite{Clarke:1998:CT,Clarke:1998:CTKF,Kupfer:1991:KF,Saed:2017:etal,Traugutt:2017:etal,Uruyama:2009:etal}. Depending on the fabrication  process, polydomains may have very different material properties and behaviors. Monodomains, where mesogen molecules are uniformly aligned throughout the material, can be formed from polydomains through mechanical stretching or by cooling an isotropic material under an external stress field  to reach the nematic phase.
 
 An ideal continuum model for monodomains is provided by the neoclassical strain-energy function proposed in \cite{Bladon:1994:BTW,Warner:1988:WGV,Warner:1991:WW}. This is a phenomenological model based on the molecular network theory of rubber elasticity \cite{Treloar:2005}. The parameters appearing in the neo-Hookean-type strain energy can be obtained through statistical averaging at microscopic scale or derived from macroscopic shape changes at small strain \cite{Warner:1996:WT,Warner:2007:WT}. Since elastic stresses dominate over Frank elasticity induced by the distortion of mesogens alignment \cite{Bai:2020:BB,Modes:2010:MBW,Modes:2011:MBW}, Frank effects are generally ignored \cite{deGennes:1993:dGP,Frank:1958}. The neoclassical formulation has been extended to polydomains by assuming that every single domain has the same strain-energy density as a monodomain \cite{Biggins:2009:BW,Biggins:2012:BWB}. These descriptions have been generalized to include nematic strain-energy densities based on phenomenological hyperelastic models (e.g., Mooney-Rivlin, Gent, Ogden) that better capture the nonlinear elastic behavior at large strains \cite{Agostiniani:2012:ADS,Agostiniani:2015:ADMDS,DeSimone:2009:dST} (molecular interpretations of the Mooney-Rivlin and Gent constitutive models for rubber are presented in \cite{Fried:2002,Horgan:2002:HS}). Further generalizations are proposed in \cite{Anderson:1999:ACF,Zhang:2019:etal}. 
 
 Another important characteristic of most materials is that physical properties are subject to random variations. Typically, average properties are used and any variations around the mean are neglected when computing the response of a material. However, as we show here, one can employ information theory \cite{Shannon:1948} and the maximum entropy principle \cite{Jaynes:1957a,Jaynes:1957b,Jaynes:2003} to include some stochastic variations of the material parameters, then propagate their uncertainty to output physical responses. Comprehensive reviews on the information-theoretic approach in elasticity are presented in \cite{Guilleminot:2017:GS,Guilleminot:2020,Soize:2017}. 

Our main focus is on large strain instabilities of LCE bodies acted upon by external loads. In addition to the recurring phenomenon of \emph{soft elasticity}, where alternating shear stripes develop at very low stress if a nematic body is stretched, we explore theoretically a set of classical instabilities inherited from parent polymeric networks, namely: \emph{necking} under tensile load, \emph{cavitation} of a nematic sphere where a void nucleates at its center when uniform tensile traction is imposed, and \emph{inflation instability} of an internally pressurized shell where the pressure increases, decreases, and then increases again. The aim is to determine conditions for the onset of instability and show how nematic materials perform compared to their purely elastic analogue. Moreover, for these problems, the propagation of stochastic variation from input material parameters to output mechanical behavior is mathematically traceable, making our stochastic approach both mathematically and mechanically transparent. The main effect of random variations is to replace a well-defined bifurcation point by a probability that a material will undergo an instability as a function of the bifurcation parameter. Such fundamental problems are important in their own right and may stimulate related mechanical testing of  nematic materials. 

\section{General set-up}
 
Following the classical work of Flory \cite{Flory:1961} on polymer elasticity, we use the stress-free state of a virtual isotropic phase at high temperature as the reference configuration \cite{Bai:2020:BB,Cirak:2014:CLBW,DeSimone:1999,DeSimone:2000:dSD,DeSimone:2002:dSD,DeSimone:2009:dST}, rather than the nematic phase in which the cross-linking might have been produced \cite{Anderson:1999:ACF,Bladon:1994:BTW,Verwey:1996:VWT,Warner:1994:WBT,Warner:1988:WGV,Warner:1991:WW,Zhang:2019:etal}. Within this theoretical framework, the material deformation due to the interaction between external stimuli and mechanical loads can be expressed as a composite deformation from a reference configuration to the current configuration, via an elastic deformation followed by a natural (stress free) shape change. The multiplicative decomposition of the associated gradient tensor is similar in some respects to those found in the constitutive theories of thermoelasticity, elastoplasticity, and growth \cite{goriely17,Lubarda:2004} (see also \cite{Goodbrake:2021:GGY,Sadik:2017:SY}), but is different on one major aspect, namely: the stress-free geometrical change is superposed on the elastic deformation, which is applied directly to the reference state. This difference is important since, although the elastic configuration obtained by this deformation may not be observed in practice, it may still be possible for the nematic body to assume such a configuration under suitable external stimuli. The resulting elastic stresses can then be used to analyze the final deformation where the particular geometry also plays a role \cite{Mihai:2020b:MG,Mihai:2020c:MG}. 
 
\begin{figure}[htbp]
	\begin{center}
		\includegraphics[width=0.6\textwidth]{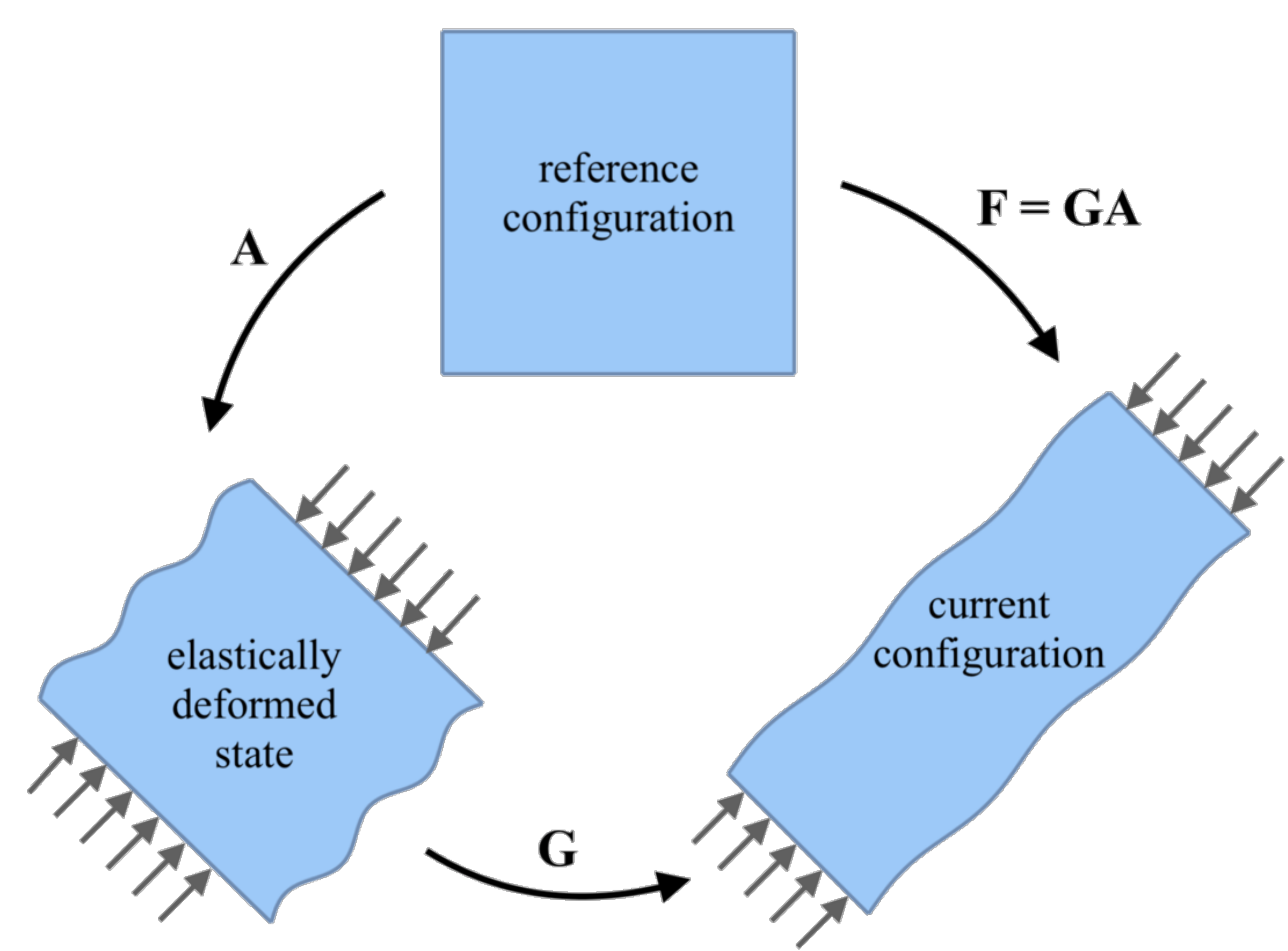}
		\caption{Multiplicative decomposition of the deformation gradient for a nematic elastomer. Unlike traditional theories of anelasticity, the elastic deformation is applied first to the reference configuration and then followed by an anelastic strain deformation.}\label{NLC:fig:FGA-diagram}
	\end{center}
\end{figure}

To describe an incompressible nematic material, we combine isotropic hyperelastic and neoclassical strain-energy density functions as follows \cite{Mihai:2020a:MG},
\begin{equation}\label{NLC:eq:W:Fn}
\overline{W}^{(nc)}(\textbf{F},\textbf{n})=W\left(\textbf{F}\textbf{G}_{0}^{-1}\right)+W^{(nc)}(\textbf{F},\textbf{n}),
\end{equation}
where, on the right-hand side, the first term is the energy of the `parent' elastic matrix, and the second term is the neoclassical-type function. Specifically: $\textbf{n}$ is a unit vector for the localized direction of uniaxial nematic alignment in the present configuration; $\textbf{F}=\textbf{G}\textbf{A}$ is the deformation gradient tensor with respect to the reference isotropic state (see Figure~\ref{NLC:fig:FGA-diagram} and also Figure~1 of \cite{DeSimone:2009:dST}), with $\textbf{G}=a^{1/3}\textbf{n}\otimes\textbf{n}+a^{-1/6}\left(\textbf{I}-\textbf{n}\otimes\textbf{n}\right)$ the  `spontaneous' (or `natural') deformation tensor and $\textbf{A}$ the (local) elastic deformation tensor; $\textbf{G}_{0}=a^{1/3}\textbf{n}_{0}\otimes\textbf{n}_{0}+a^{-1/6}\left(\textbf{I}-\textbf{n}_{0}\otimes\textbf{n}_{0}\right)$ is the spontaneous deformation tensor with $\textbf{n}_{0}$ the director orientation at cross-linking (which may be spatially varying); and $a>0$ is a temperature-dependent shape parameter, which we assume to be spatially independent (i.e., no differential swelling). We denote by $\otimes$ the usual tensor product of two vectors, and by $\textbf{I}=\text{diag}(1,1,1)$ the identity second order tensor. 

Recognizing that some uncertainties may arise in the mechanical responses of liquid crystal elastomers, and inspired by recent developments in stochastic finite elasticity \cite{Fitt:2019:FWWM,Mihai:2021:MA,Mihai:2019a:MDWG,Mihai:2019b:MDWG,Mihai:2019c:MDWG,Mihai:2018:MWG,Mihai:2019a:MWG,Mihai:2019b:MWG,Mihai:2020:MWG,Staber:2015:SG,Staber:2016:SG,Staber:2017:SG,Staber:2018:SG,Staber:2019:SGSMI}, we assume that the model parameters are defined as random variables drawn from given probability distributions. In practice, material parameters take on different values, corresponding to possible outcomes of experimental tests. The maximum entropy principle then allows us to explicitly construct prior probability laws for the model parameters, given the available information. Explicit derivations of probability distributions for the elastic parameters of stochastic homogeneous isotropic hyperelastic models calibrated to experimental data for rubber-like material and soft tissues were presented in \cite{Fitt:2019:FWWM,Mihai:2018:MWG,Staber:2017:SG}. Intuitively, such a stochastic body can be regarded as an ensemble (or population) of bodies that are equal in size and have the same geometry, and each body in the ensemble is made from a single homogeneous material with the  parameters not known with certainty, but distributed according to probability density functions that are calibrated to macroscopic experimental measurements. These models reduce to the usual deterministic ones when the parameters are single-valued constants. 

\section{Soft elasticity and stress plateaus}\label{NLC:sec:striping}

Many macroscopic deformations of nematic liquid crystal elastomers induce a re-orientation of the director with a general tendency for the director to become parallel to the direction of the largest principal stretch. This re-orientation is  typically uniform across the material. However, non-uniform behaviors are also possible. In particular, under appropriate uniaxial tension or biaxial stretch, bifurcation to a pattern of stripe domains is generated, where adjacent stripes deform by the same shear but in opposite directions. Early experimental investigations of this phenomenon, known as \emph{soft elasticity} \cite{Olmsted:1994,Verwey:1995:VW,Warner:1994:WBT}, were reported in \cite{Finkelmann:1997:FKTW,Kundler:1995:KF,Talroze:1999:etal,Zubarev:1999:etal}. Its theoretical explanation is that, for these materials, the energy is minimized by passing through a state exhibiting a microstructure of many homogeneously deformed parts \cite{Conti:2002:CdSD,DeSimone:2000:dSD,DeSimone:2009:dST,Fried:2004:FS,Fried:2005:FS,Fried:2006:FS,Kundler:1995:KF}. A natural question is then: \emph{How does soft elasticity depend on the material parameters?} 

\begin{figure}[htbp]
	\begin{center}
		\subfigure[]{\includegraphics[width=0.49\textwidth]{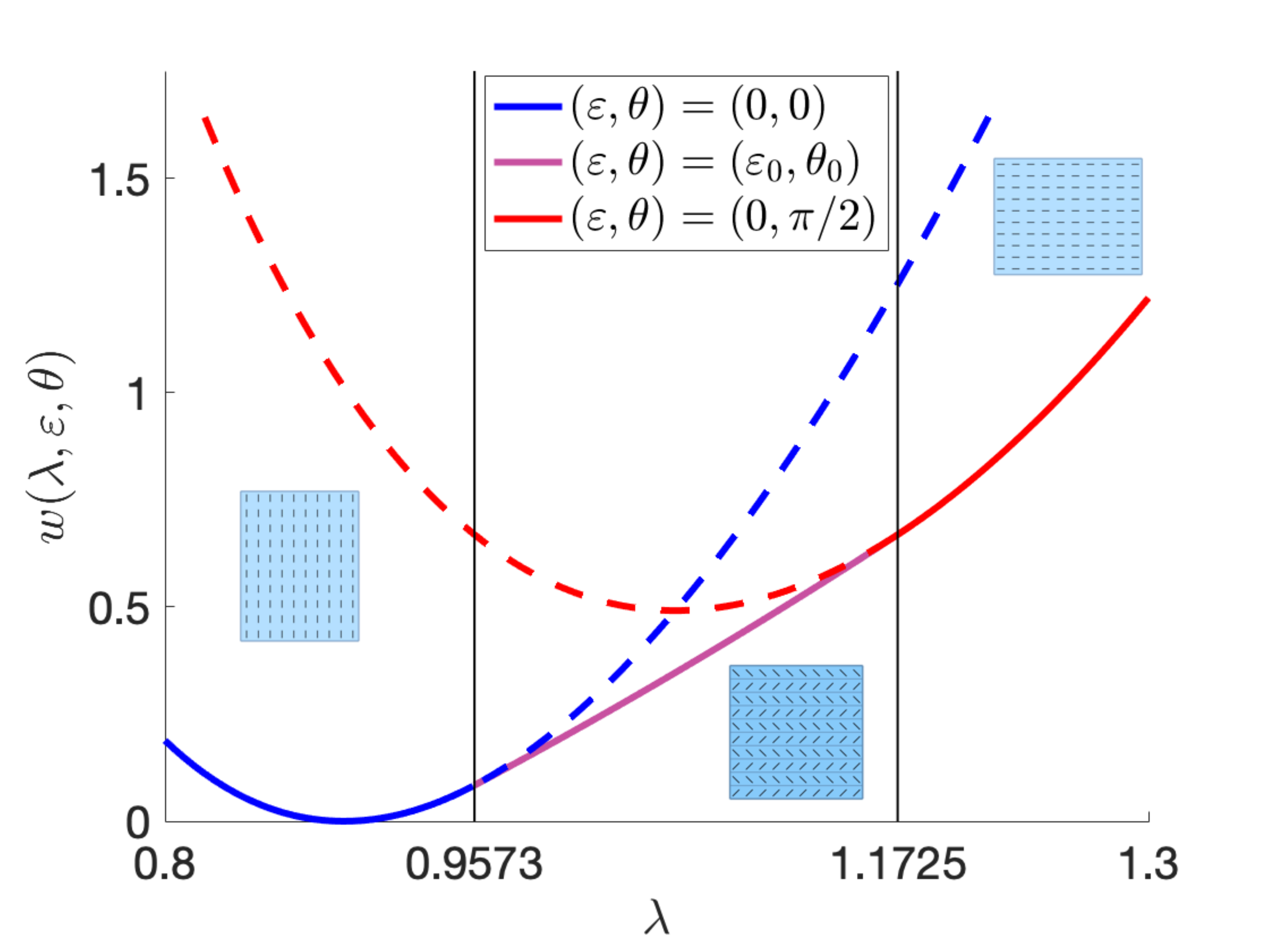}}
		\subfigure[]{\includegraphics[width=0.49\textwidth]{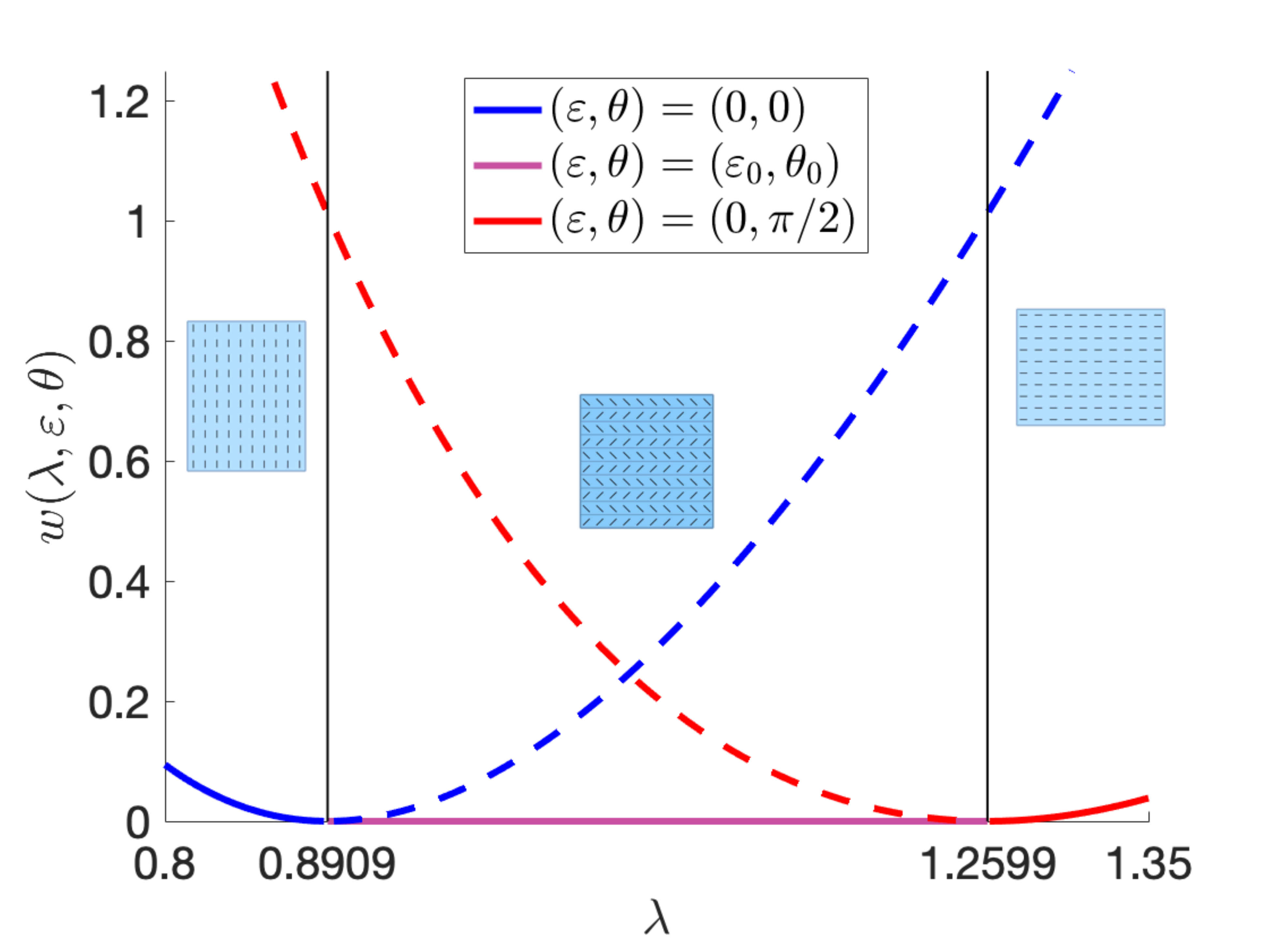}}\\
		\subfigure[]{\includegraphics[width=0.49\textwidth]{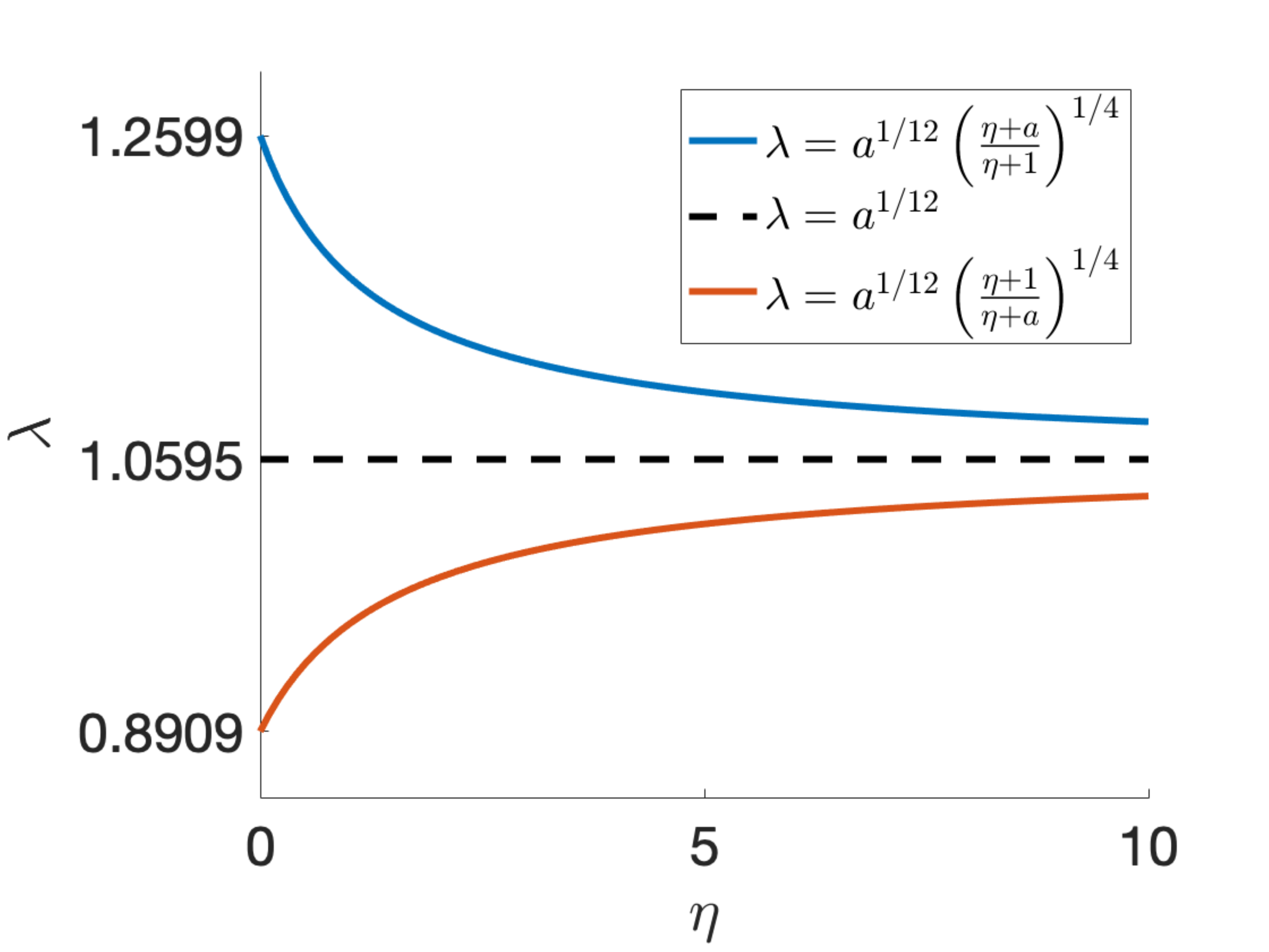}}
		\subfigure[]{\includegraphics[width=0.49\textwidth]{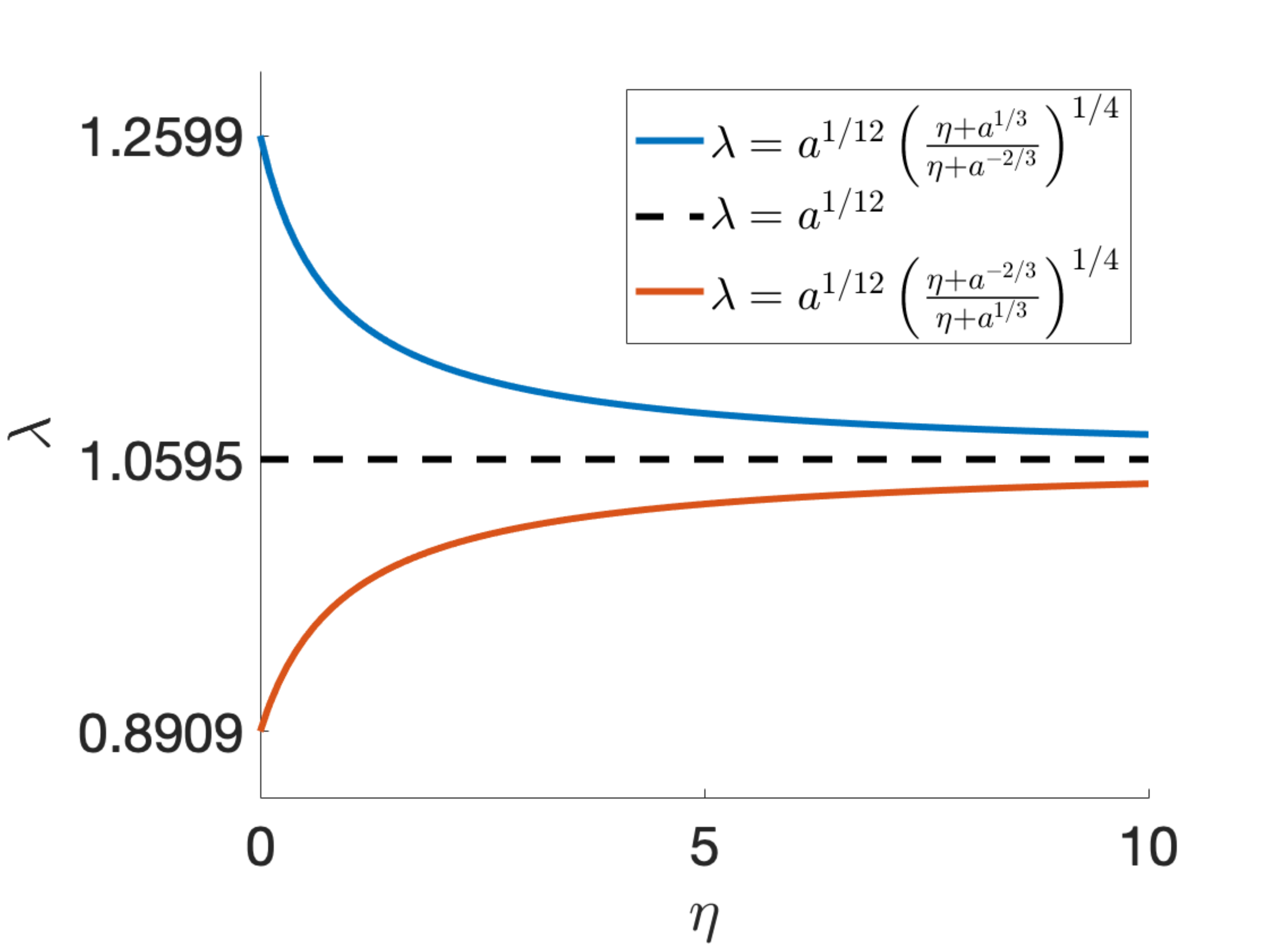}}
		\caption{(a)-(b) The strain-energy function $w(\lambda,\varepsilon,\theta)$, for $(\varepsilon,\theta)=(0,0)$, $(\varepsilon,\theta)=(\varepsilon_{0},\theta_{0})$, with $\varepsilon_{0}$ and $\theta_{0}$ given by \eqref{NLC:eq:epstheta0}, and $(\varepsilon,\theta)=(0,\pi/2)$, when $a=2$ and (a) $\mu^{(1)}=4.05$ or (b) $\mu^{(1)}=0$, while $\mu^{(2)}=4.05$. The two vertical lines correspond to the lower and upper bounds on $\lambda$, given by equations \eqref{NLC:eq:bounds1}  and \eqref{NLC:eq:bounds2}, respectively. Between these bounds, the second solution, with $(\varepsilon,\theta)=(\varepsilon_{0},\theta_{0})$, minimizes the energy. The lower and upper bounds on $\lambda$, given by (c) \eqref{NLC:eq:bounds1}-\eqref{NLC:eq:bounds2}, and (d) \eqref{NLC:eq:bounds3}-\eqref{NLC:eq:bounds4}, respectively, as functions of the parameter ratio $\eta=\mu^{(1)}/\mu^{(2)}$ when $a=2$. For $\eta\to\infty$, the model approaches a purely elastic form where the homogeneous deformation is always stable. For $\eta=0$, the bounds are the same, and correspond to the neoclassical form.}\label{NLC:fig:energybounds}
	\end{center}
\end{figure}

\begin{figure}[htbp]
	\begin{center}
		\includegraphics[width=0.49\textwidth]{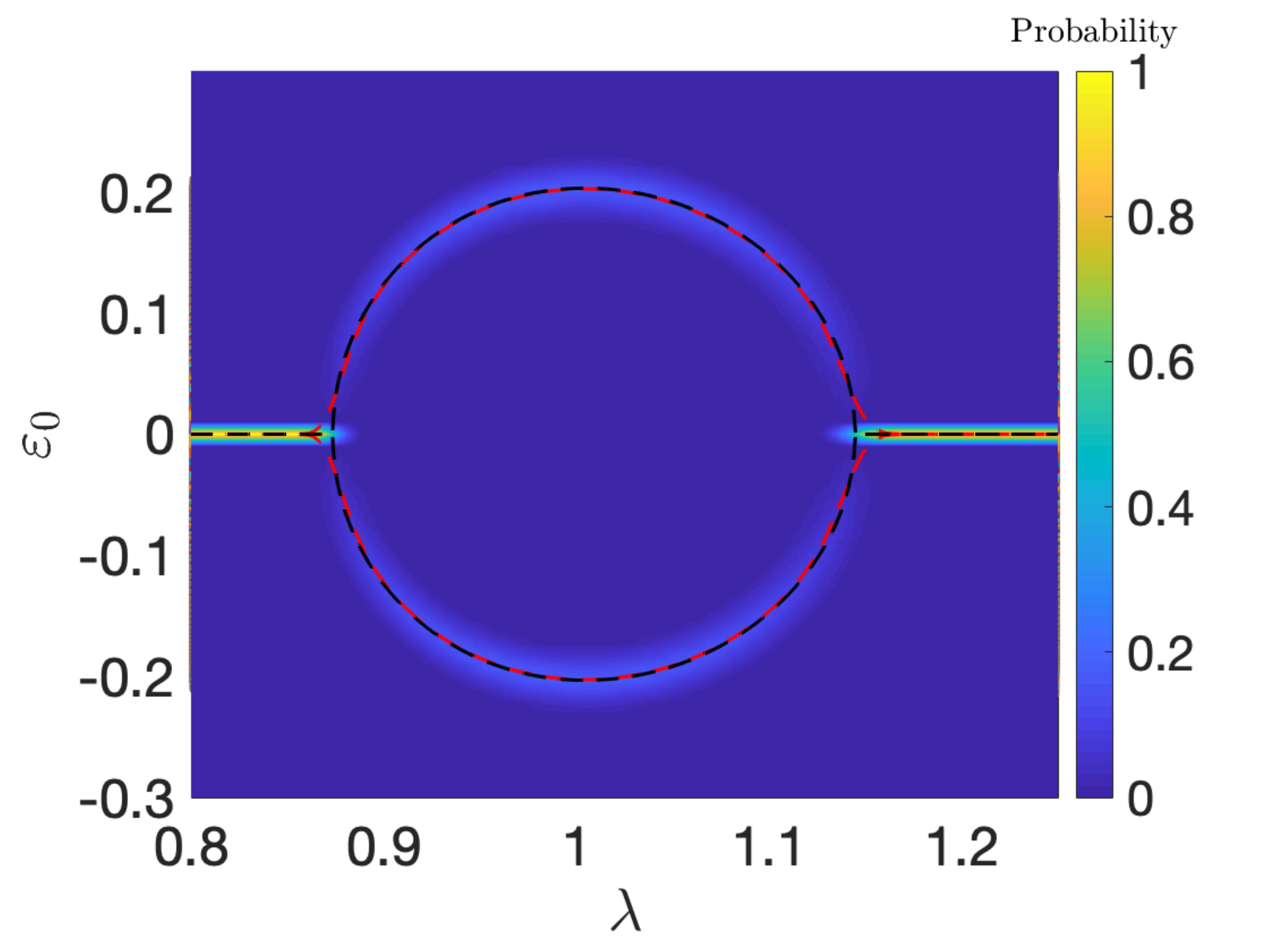}
		\includegraphics[width=0.49\textwidth]{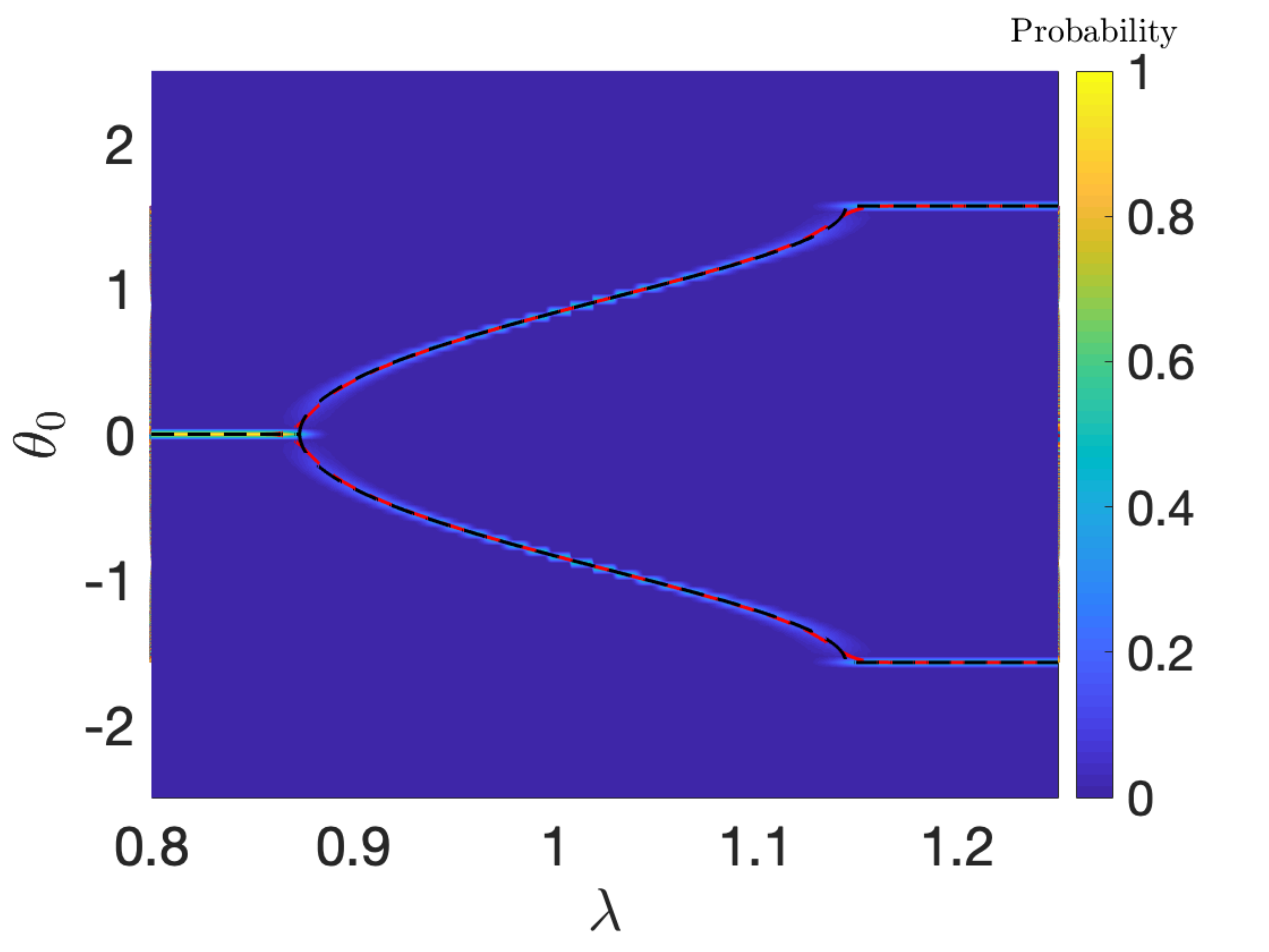}\\
		\includegraphics[width=0.49\textwidth]{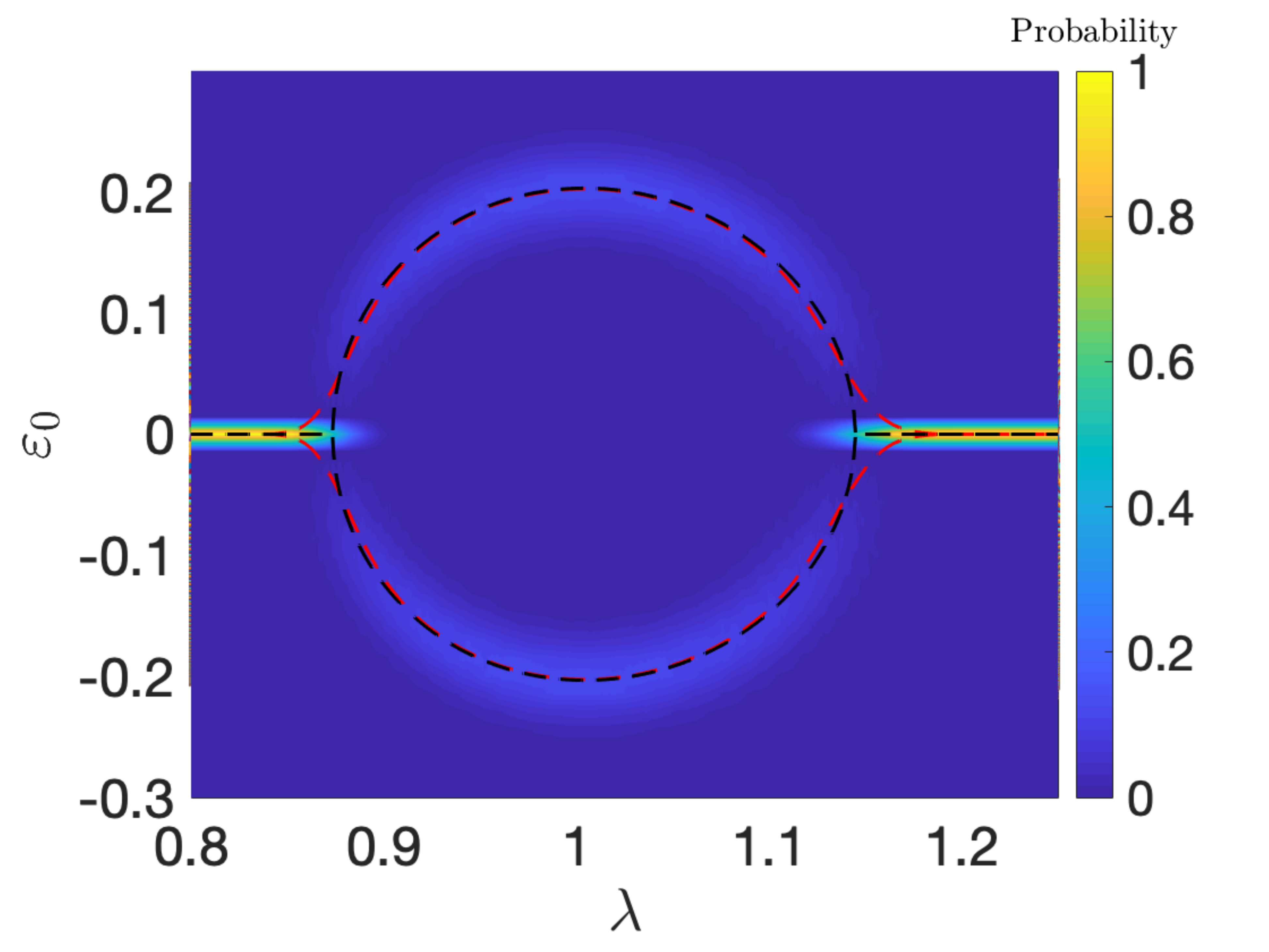}
		\includegraphics[width=0.49\textwidth]{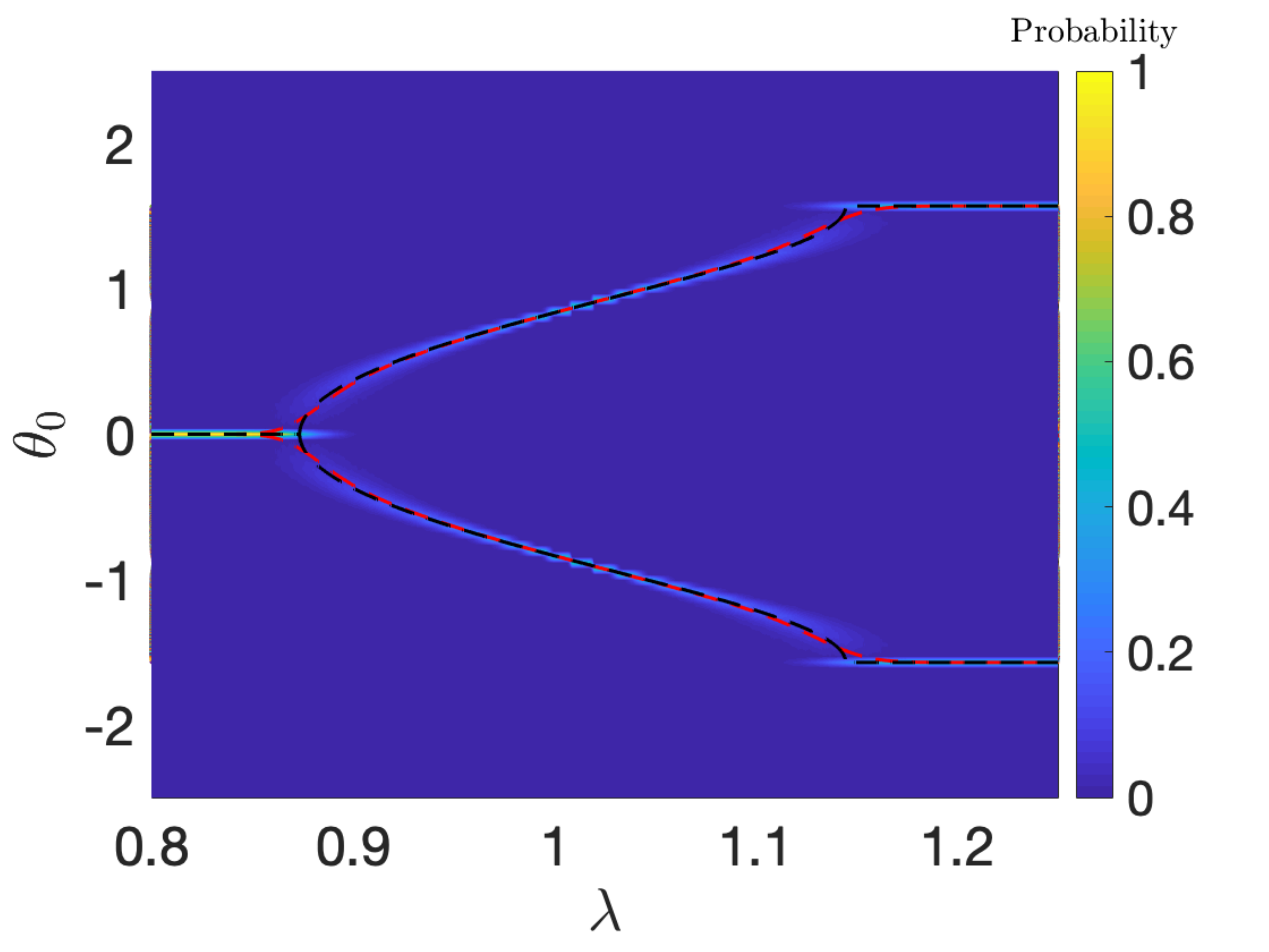}
		\caption{The stochastic shear parameter $\varepsilon_{0}$ and director angle $\theta_{0}$, given by \eqref{NLC:eq:epstheta0}, when: (a)-(b) $a=2$, while the shear modulus $\mu$ follows a Gamma distribution with shape and scale parameters $\rho_{1}=405$ and $\rho_{2}=0.01$, respectively, and $R^{(1)}=\mu^{(1)}/\mu$ is drawn from a Beta distribution with hyperparameters $\xi_{1}=100$ and $\xi_{2}=100$; (c)-(d) $\mu=4.05$ and $R^{(1)}=\mu^{(1)}/\mu=0.5$, while the shape parameter $a$ follows a Gamma distribution with $\rho_{1}=200$, $\rho_{2}=0.01$. In (a)-(b), the dashed black lines correspond to the deterministic solutions based only on the mean values $\underline{\mu}=\rho_{1}\rho_{2}=4.05$ and $\underline{R}_{1}=\xi_{1}/(\xi_{1}+\xi_{2})=0.5$, whereas the red versions show the arithmetic mean value solutions. In (c)-(d), the dashed black lines correspond to the deterministic solutions based only on the mean value $\underline{a}=\rho_{1}\rho_{2}=2$, whereas the red versions show the arithmetic mean value solutions. In this case, there is a significant difference between the mean value solutions and the deterministic solutions.}\label{NLC:fig:muastoch}
	\end{center}
\end{figure}

For simplicity, we select an incompressible neo-Hookean-type strain-energy function \cite{Treloar:1944}, 
\begin{equation}\label{NLC:eq:W:NH}
W(\textbf{A})=\frac{\mu^{(1)}}{2}\left[\text{tr}\left(\textbf{A}\textbf{A}^\text{T}\right)-3\right],
\end{equation}
where the superscript ``T'' represents the transpose operator, ``tr'' denotes the trace operator, and $\mu^{(1)}\geq0$ is constant, together with the neoclassical strain-energy function \cite{Cirak:2014:CLBW,DeSimone:2000:dSD,DeSimone:2009:dST},
\begin{equation}\label{NLC:eq:Wnc:NH}
W^{(nc)}(\textbf{F},\textbf{n})=\frac{\mu^{(2)}}{2}\left\{a^{1/3}\left[\text{tr}\left(\textbf{F}\textbf{F}^\text{T}\right)-\left(1- a^{-1}\right)\textbf{n}\cdot\textbf{F}\textbf{F}^\text{T}\textbf{n}\right]-3\right\},
\end{equation}
with $\mu^{(2)}\geq0$ constant. For the composite model function defined by \eqref{NLC:eq:W:Fn}, the shear modulus at infinitesimal strain is $\mu=\mu^{(1)}+\mu^{(2)}>0$ \cite{Mihai:2017:MG}. However, our results can be easily extended to other choice of strain-energy density functions \cite{Mihai:2020a:MG}.

We analyze shear striping under biaxial stretch, and assume that the nematic director can only rotate in the biaxial plane. To achieve this, we set $\textbf{n}_{0}=[0,0,1]^\text{T}$ and  $\textbf{n}=\left[0,\sin\theta,\cos\theta\right]^\text{T}$, where $\theta\in[0,\pi/2]$, in a Cartesian system of reference, and examine small shear perturbations of biaxial extensions, with gradient tensor \cite{Mihai:2020a:MG}
\begin{equation}\label{NLC:eq:F:stretch}
\textbf{F}=
\left[
\begin{array}{ccc}
a^{-1/6} & 0 & 0\\
0 & \lambda & \varepsilon\\
0 & 0 & a^{1/6}\lambda^{-1}
\end{array}
\right],
\end{equation}
where $a>1$ is the nematic shape parameter, $\lambda>0$ is the stretch ratio, and $\varepsilon>0$ is the small perturbation. Denoting $w(\lambda,\varepsilon,\theta)=\overline{W}^{(nc)}(\textbf{F},\textbf{n})$, for sufficiently small values of $\varepsilon$ and $\theta$, we find that, when $\mu^{(1)}>0$ and $\mu^{(2)}=0$ (purely elastic case), the equilibrium state with $\varepsilon=0$ and $\theta=0$ is stable. If $\mu^{(2)}>0$, then this equilibrium state is unstable for 
\begin{equation}\label{NLC:eq:bounds1}
a^{-1/6}\leq a^{1/12}\left(\frac{\eta+1}{\eta+a}\right)^{1/4}<\lambda<a^{1/12},
\end{equation}
where the \emph{elasto-nematic ratio} $\eta=\mu^{(1)}/\mu^{(2)}$ gives the relative magnitude of the elastic and neoclassical contributions. There is also an equilibrium state with  $\varepsilon=0$ and $\theta=\pi/2$, where the nematic director is fully rotated so that it aligns uniformly with the direction of macroscopic extension. By symmetry arguments, this state is unstable for
\begin{equation}\label{NLC:eq:bounds2}
a^{1/12}<\lambda<a^{1/12}\left(\frac{\eta+a}{\eta+1}\right)^{1/4}\leq a^{1/3}.
\end{equation}
When $\lambda$ satisfies \eqref{NLC:eq:bounds1}  or \eqref{NLC:eq:bounds2}, we obtain
\begin{equation}\label{NLC:eq:epstheta0}
\varepsilon_{0}=\pm\frac{\lambda(a-1)\sin(2\theta_{0})}{2\sqrt{\left(\eta+1\right)\left(\eta+a\right)}},\qquad
\theta_{0}=\pm\arccos\sqrt{\frac{a^{1/6}\sqrt{\left(\eta+1\right)\left(\eta+a\right)}}{\lambda^2(a-1)}-\frac{\eta+1}{a-1}}.
\end{equation}
For the resulting strip pattern, the gradient tensors of alternating shear deformations in two adjacent stripe domains are $\textbf{F}_{\pm}$ with $\varepsilon=\pm\varepsilon_{0}$, respectively. The two deformations are geometrically compatible in the sense that there exist two non-zero vectors $\textbf{q}$ and $\textbf{p}$, such that the Hadamard jump condition $\textbf{F}_{+}-\textbf{F}_{-}=\textbf{q}\otimes\textbf{p}$ is satisfied, 
where $\textbf{p}$ is the normal vector to the interface between the two phases corresponding to the deformation gradients $\textbf{F}_{+}$ and $\textbf{F}_{-}$ \cite{Ball:1987:BJ,Ball:1992:BJ,Mihai:2017a:MN,Mihai:2017b:MN}. In other words, $\textbf{F}_{+}$ and $\textbf{F}_{-}$ are rank-one connected, i.e., $\mathrm{rank}\left(\textbf{F}_{+}-\textbf{F}_{-}\right)=1$. If $\eta=0$, then the above equilibrium states are unstable for $\lambda\in\left(a^{-1/6},a^{1/12}\right)$ and $\lambda\in\left(a^{1/12},a^{1/3}\right)$, respectively. Thus, soft elasticity is always presented by the purely neoclassical model \cite{Conti:2002:CdSD,DeSimone:2009:dST}. When $\eta\to\infty$, there is no shear striping since the material is practically elastic. 

The strain-energy function $w(\lambda,\varepsilon,\theta)$ is illustrated in Figure~\ref{NLC:fig:energybounds}(a)-(b). Note that, for $\lambda$ with values between the lower and upper bounds given by \eqref{NLC:eq:bounds1} and \eqref{NLC:eq:bounds2}, respectively, the minimum energy is attained for $(\varepsilon,\theta)=(\varepsilon_{0},\theta_{0})$. Assuming that loading is applied in the second direction, the first Piola-Kirchhoff axial stress in this direction is equal to $P^{(nc)}_{2}=\mathrm{d}w(\lambda,\varepsilon,\theta)/\mathrm{d}\lambda$. Figure~\ref{NLC:fig:energybounds}(b) then suggests that, when $\mu^{(1)}=0$, i.e., for the purely neoclassical form, the director rotates and alternating shear stripes develop for $\lambda\in\left(a^{-1/6}, a^{1/3}\right)$, at zero load, since the slope of the curve is equal to zero within this interval. In contrast, if $\mu^{(1)}>0$, then from Figure~\ref{NLC:fig:energybounds}(a), we infer that the applied load increases with deformation and is almost constant but non-zero while the director rotates.

Due to the geometric compatibility, and since the intervals for stretch ratios $\lambda$ where shear striping occurs are at a maximum length when $\textbf{n}_{0}=[0,0,1]^\text{T}$, the bounds \eqref{NLC:eq:bounds1}-\eqref{NLC:eq:bounds2} give also the maximum interval for shear striping when $\textbf{n}_{0}$ is not uniformly aligned. The minimum length of those intervals is attained for monodomains with $\textbf{n}_{0}=[0,1,0]^\text{T}$. Experimental results for monodomains where the tensile load forms different angles with the initial nematic director are reported in \cite{Mistry:2019:MG,Okamoto:2021:OSU}. If $\textbf{G}_{0}=\textbf{I}$ (see \cite{Biggins:2009:BW}), then the solution with $\varepsilon=0$ and $\theta=0$ is unstable for 
\begin{equation}\label{NLC:eq:bounds3}
a^{-1/6}\leq a^{1/12}\left(\frac{\eta+a^{-2/3}}{\eta+a^{1/3}}\right)^{1/4}<\lambda<a^{1/12},
\end{equation}
and that with $\varepsilon=0$ and $\theta=\pi/2$ is unstable for 
\begin{equation}\label{NLC:eq:bounds4}
a^{1/12}<\lambda<a^{1/12}\left(\frac{\eta+a^{1/3}}{\eta+a^{-2/3}}\right)^{1/4}\leq a^{1/3}.
\end{equation}
For example, when $a=2$, the bounds given by \eqref{NLC:eq:bounds1}-\eqref{NLC:eq:bounds2}, and by \eqref{NLC:eq:bounds3}-\eqref{NLC:eq:bounds4}, respectively, are plotted as functions of the parameter ratio $\eta$ in Figure~\ref{NLC:fig:energybounds}(c)-(d).

In addition, when the model parameters are random variables, characterized by probability distributions, we are interested in the probability that shear striping occurs under a given stretch. Here, we show the effects of fluctuations in the shear modulus $\mu$ and the shape parameter $a$ separately, i.e., we first set $\mu$ as a random variable while $a$ is a single-valued constant, then keep $\mu$ constant and let $a$ fluctuate. In Figure~\ref{NLC:fig:muastoch}(a)-(b), the stochastic shear parameter $\varepsilon_{0}$ and director angle $\theta_{0}$, given by \eqref{NLC:eq:epstheta0}, are represented when the stochastic material parameters were chosen such that their mean values are the same as the deterministic values in Figure~\ref{NLC:fig:energybounds}(a). For both stochastic versions, to increase the probability of homogeneous deformation, one must increase the value of $\eta$, whereas shear striping is certain only if the model reduces to the neoclassical one (i.e., when $\eta=0$). However, while $\eta>0$, the inherent variability in the probabilistic system means that there will always be competition between the homogeneous and inhomogeneous deformations. 

\section{Cavitation}

Cavitation in solids is the formation of a void within a solid under tensile loads, by analogy to the similar phenomenon observed in fluids. For rubber-like materials, following early studies of damage under loads \cite{Busse:1938,Yerzley:1939}, this phenomenon was first reported in \cite{Gent:1959:GL} where experiments showed that rubber cylinders ruptured under relatively small tensile dead loads by opening an internal cavity. The nonlinear elastic analysis in \cite{Ball:1982} provided the first systematic theory for the formation of a spherical cavity at the center of a sphere of isotropic incompressible hyperelastic material under a radially symmetric tensile load. Mathematically, cavitation was treated as a bifurcation from the trivial state at a critical value of the surface traction or displacement, at which the trivial solution became unstable. This work paved the way for numerous applied and theoretical studies devoted to this inherently nonlinear mechanical effect, which is not captured by the linear elasticity theory. In addition to the well-known stable cavitation post-bifurcation at the critical dead load, such that the cavity radius monotonically increases with the applied load, it was shown  in \cite{Mihai:2019c:MDWG} that unstable (snap) cavitation is also possible for some homogeneous isotropic incompressible hyperelastic materials where Backer-Ericksen inequalities \cite{BakerEricksen:1954,Marzano:1983} hold. In \cite{Mihai:2020:MWG}, static and dynamic cavitation in radially-inhomogeneous spheres with stochastic parameters were analyzed. 

\begin{figure}[htbp]
	\begin{center}
		\subfigure[]{\includegraphics[width=0.49\textwidth]{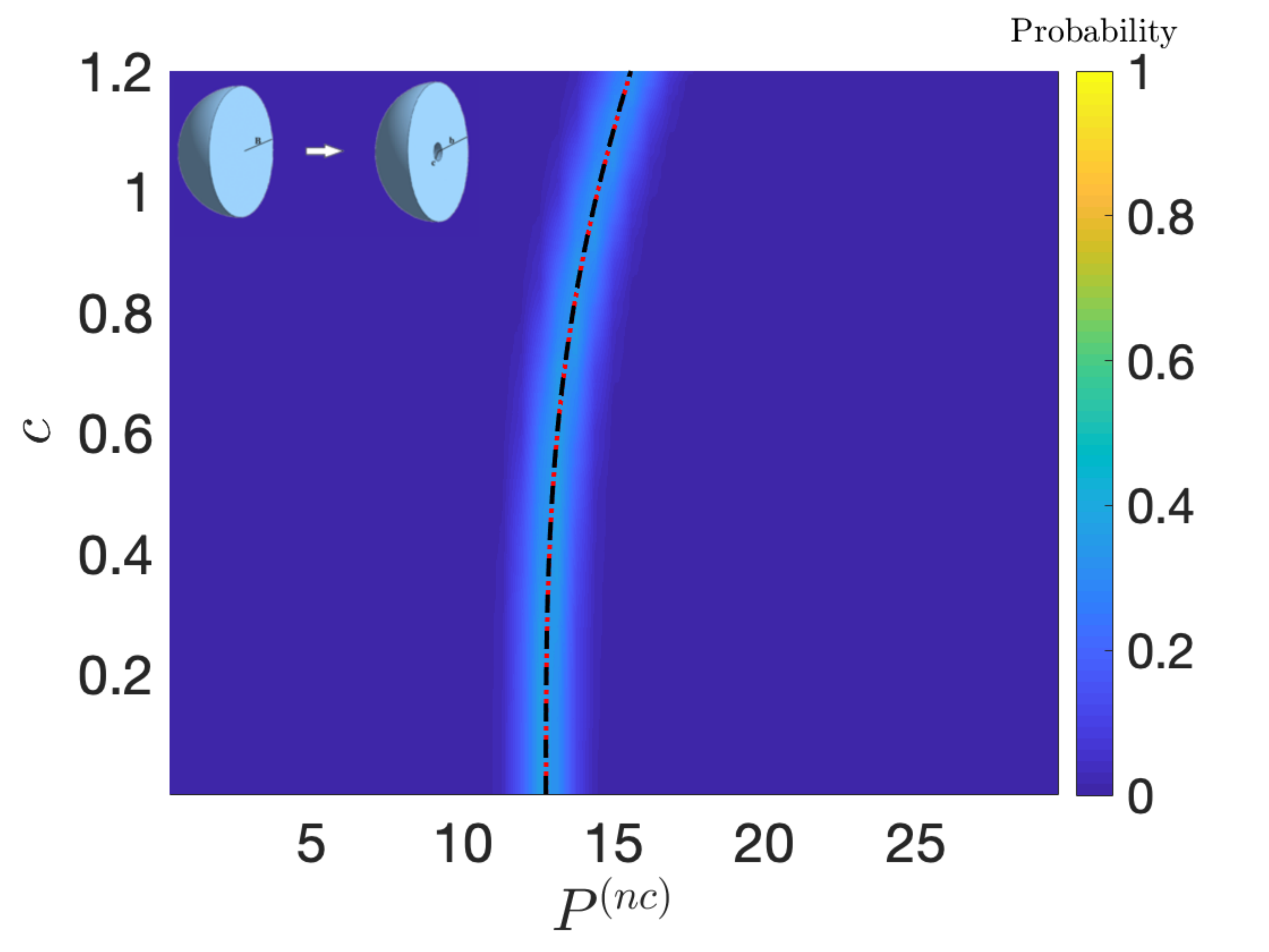}}
		\subfigure[]{\includegraphics[width=0.49\textwidth]{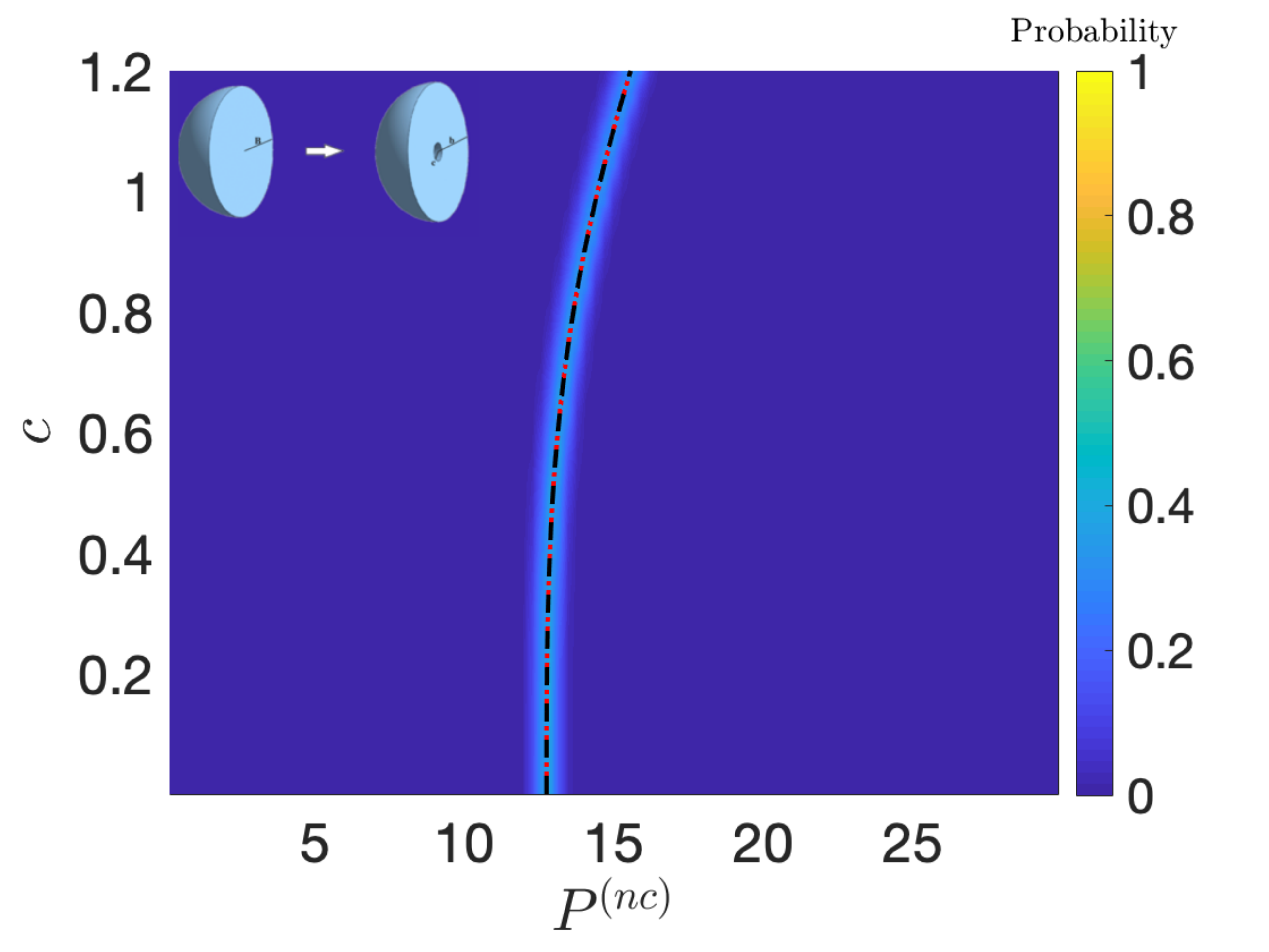}}
		\caption{Probability distribution of the applied dead-load traction causing cavitation of radius $c$ at the center of an initially unit sphere of neoclassical LCE material when: (a) $a=2$, while the shear modulus $\mu$ follows a Gamma distribution with hyperparameters $\rho_{1}=405$, $\rho_{2}=0.01$, and (b) $\mu=4.05$, while the shape parameter $a$ follows a Gamma distribution with $\rho_{1}=200$, $\rho_{2}=0.01$. The dashed black lines correspond to the deterministic solutions based only on the mean values of the parameters, whereas the red versions show the arithmetic mean value solutions. The probabilistic interval for the critical load at which the cavity nucleates is found when $c\to0$. As the bifurcation from trivial solution is supercritical, cavitation is stable, i.e., the cavity radius increases as the applied dead load increases.}\label{NLC:fig:cave-NH}
	\end{center}
\end{figure}

To study the cavitation of a nematic sphere, we consider an initially unit sphere described by the neoclassical model \eqref{NLC:eq:Wnc:NH}, where either the shear modulus $\mu$ or the nematic parameter $a$ is a random variable. Expressing all tensors in the spherical coordinates $(R,\Theta,\Phi)$ in the reference configuration, we assume that the sphere is deformed by radially symmetric inflation with deformation gradient $\textbf{F}=\mathrm{{diag}}\left(\lambda^{-2},\lambda,\lambda\right)$, while the natural deformation tensor is $\textbf{G}=\mathrm{diag}\left(a^{-1/3},a^{1/6},a^{1/6}\right)$, and $\lambda>a^{1/6}>1$. When this deformation is due to a radial dead-load traction applied uniformly on the sphere surface in the reference configuration, we are interested in the critical load that will cause a spherical cavity to open at its center. The radial traction at the outer surface necessary for an inner cavity of radius $c$ to form is illustrated in Figure~\ref{NLC:fig:cave-NH}. For the onset of cavitation, the critical load is found by letting $c\to0$. Since the bifurcation is supercritical, cavitation is stable, i.e., the cavity radius increases as the applied dead load increases. A comparison with \cite{Mihai:2020:MWG} shows that, for the nematic sphere, cavitation nucleates at a larger critical load, $P^{(nc)}_{0}$, than the corresponding  load, $P_{0}$, for the elastic sphere with the same shear modulus, and $P^{(nc)}_{0}=a^{1/3}P_{0}$. We recall that, for the neo-Hookean sphere, $P_{0}=5\mu/2$. When the LCE strain-energy function includes an additional elastic component, as in \eqref{NLC:eq:W:Fn}, so that the elasto-nematic ratio is $\eta>0$, while the shear modulus $\mu$ remains the same, the critical dead load decreases towards $P_{0}$. For example, if $\textbf{G}_{0}=\textbf{I}$, then the critical applied load is equal to $\overline{P}^{(nc)}_{0}=\left[a^{1/3}+\left(1-a^{1/3}\right)\mu^{(1)}/\mu\right]P_{0}$, and $P_{0}<\overline{P}^{(nc)}_{0}<P^{(nc)}_{0}$. 

\section{Shell inflation}

\begin{figure}[htbp]
	\begin{center}
		\subfigure[]{\includegraphics[width=0.49\textwidth]{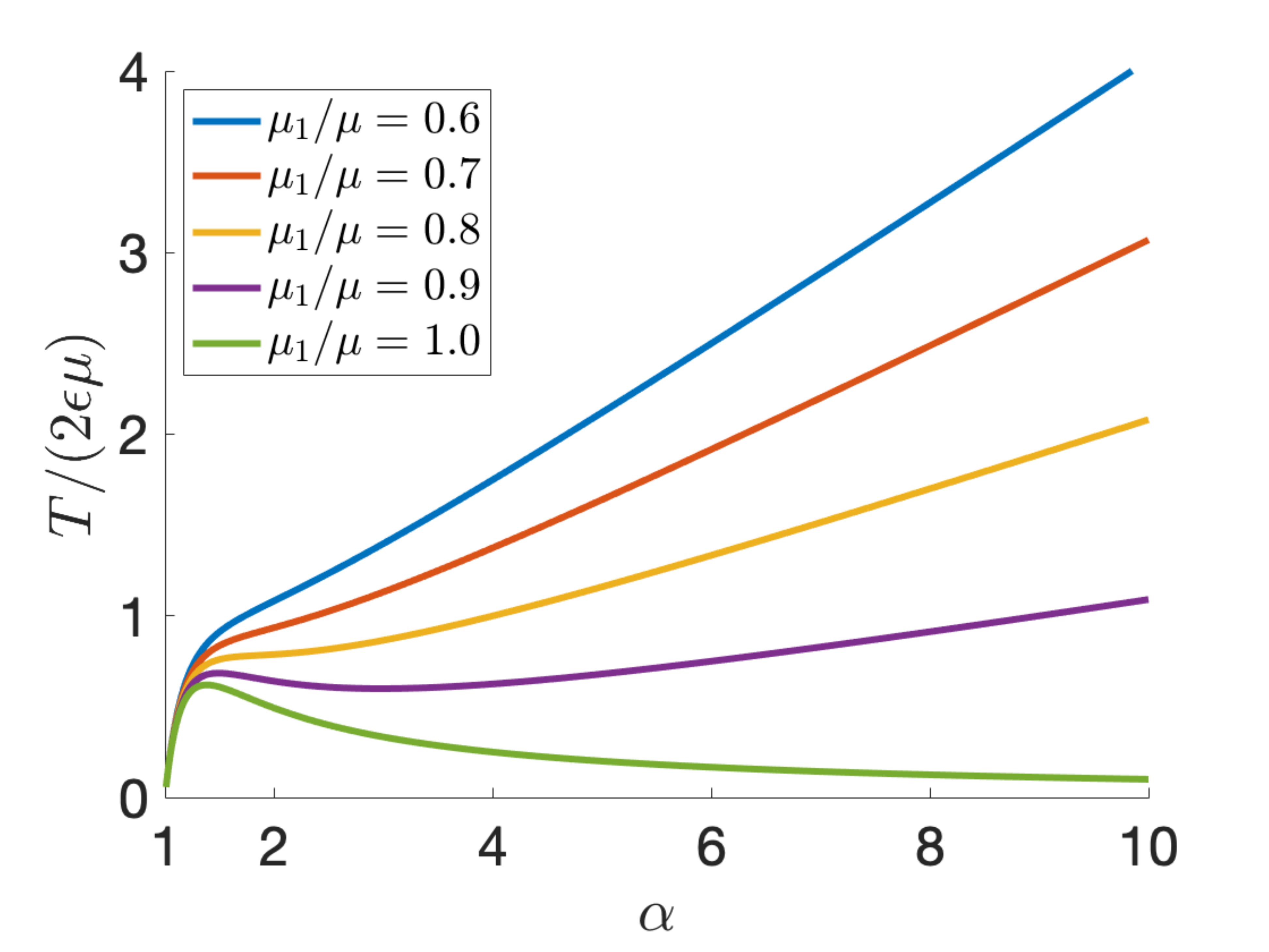}}
		\subfigure[]{\includegraphics[width=0.49\textwidth]{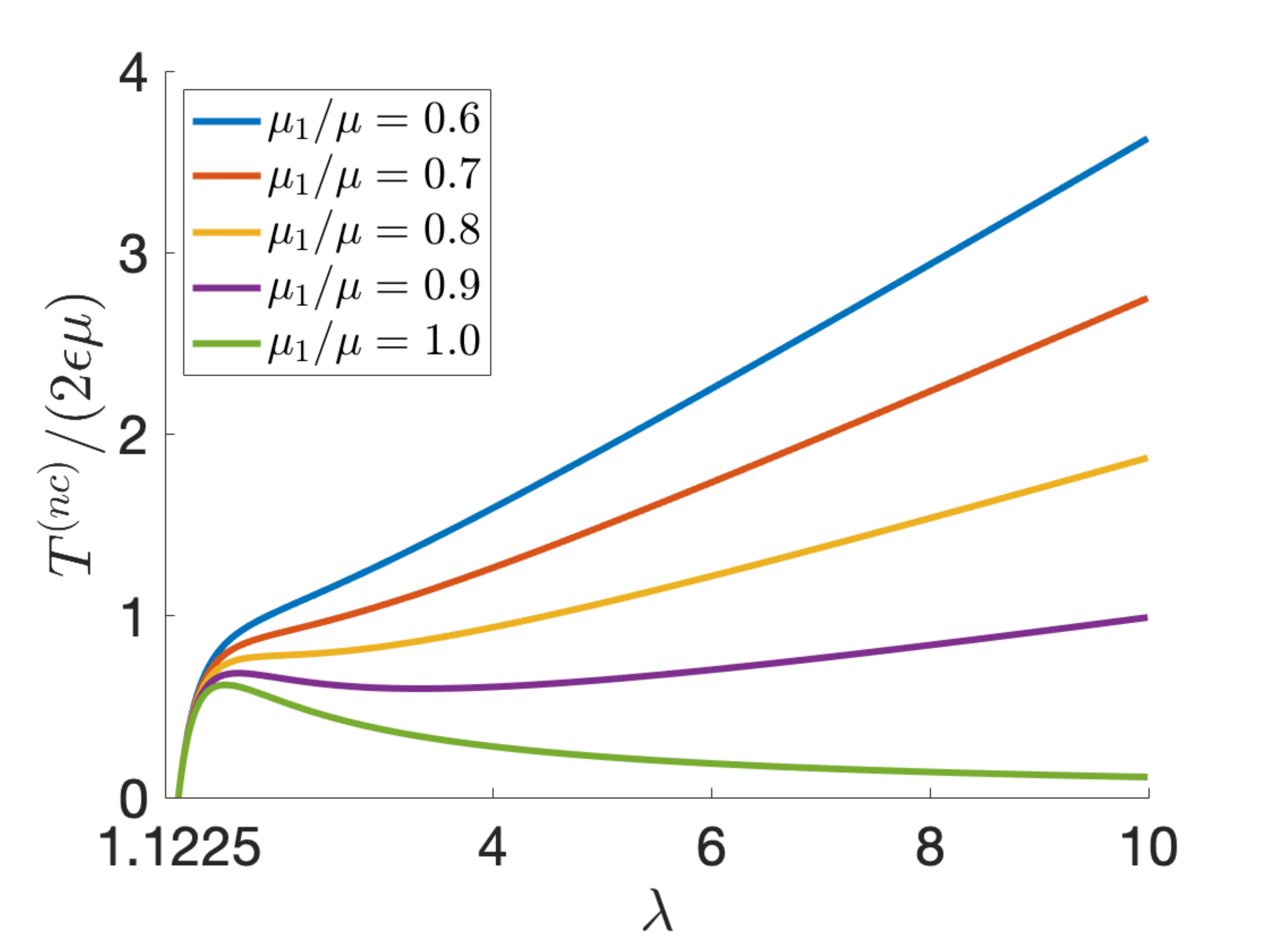}}\\
		\subfigure[]{\includegraphics[width=0.49\textwidth]{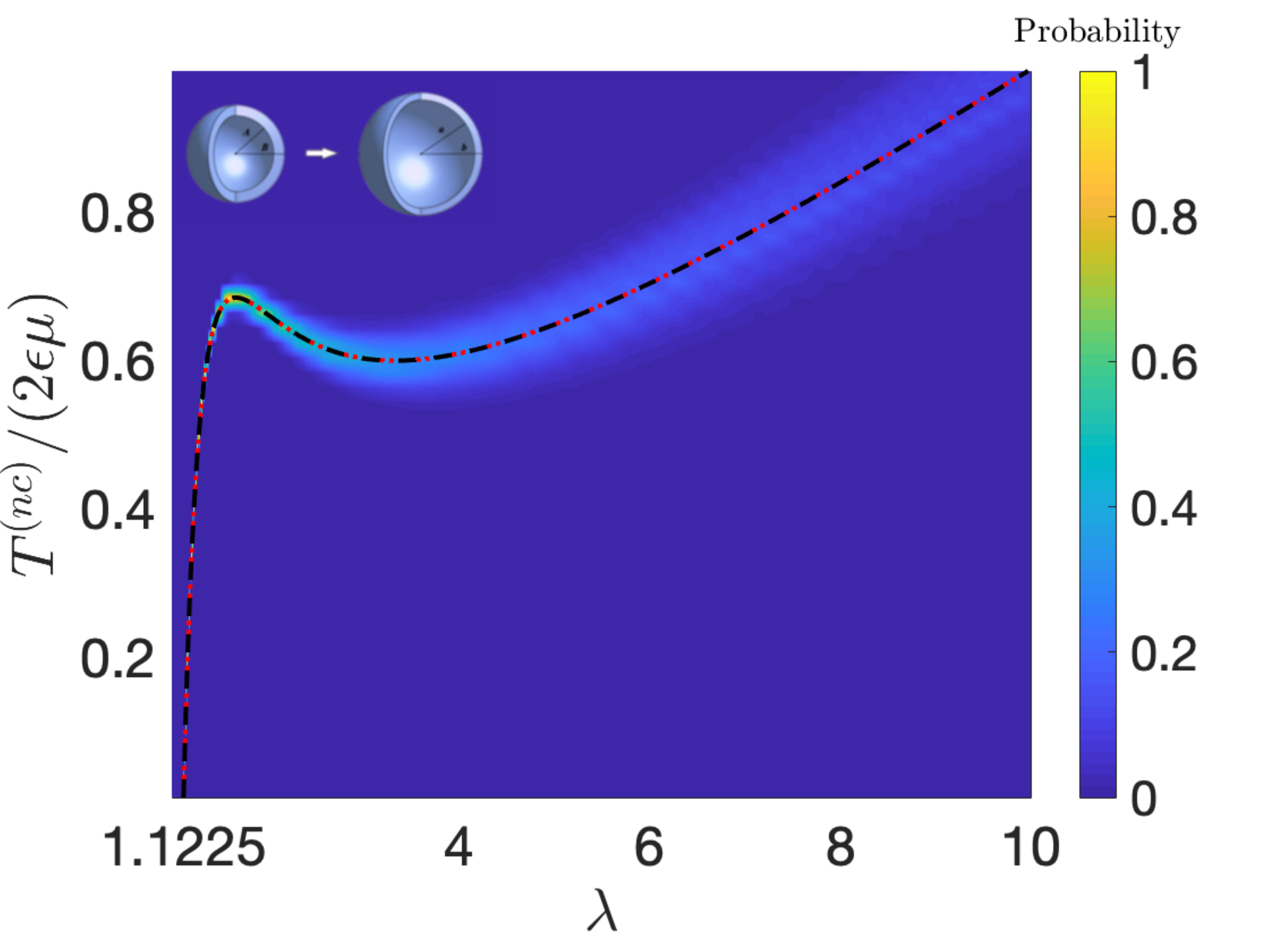}}
		\subfigure[]{\includegraphics[width=0.49\textwidth]{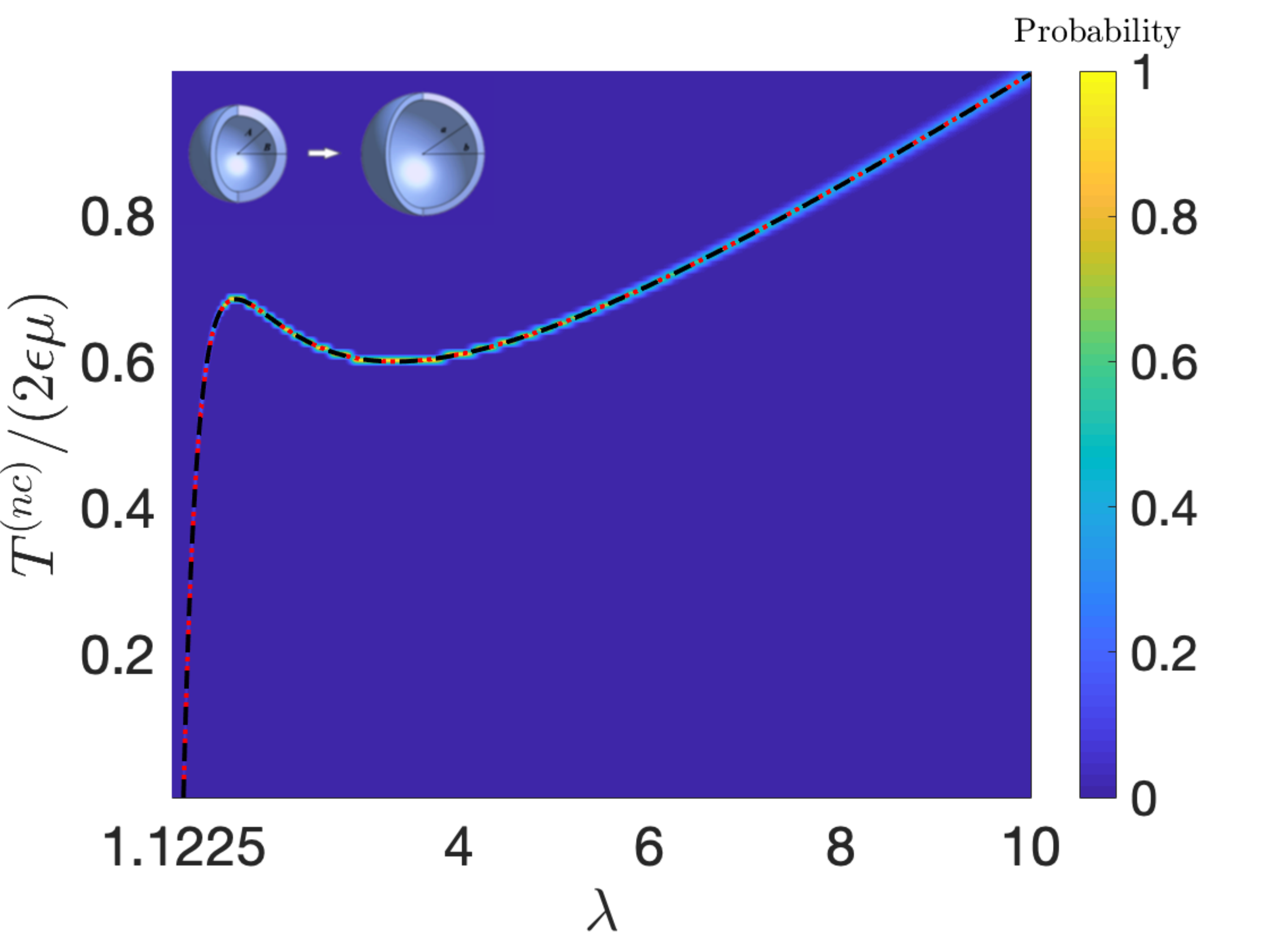}}
		\caption{The normalized internal pressure for an inflated spherical shell of: (a) Mooney-Rivlin hyperelastic material, defined by \eqref{NLC:eq:Wiso:MR}, when $\textbf{A}=\mathrm{{diag}}\left(\alpha^{-2},\alpha,\alpha\right)$, with $\alpha>1$; (b) Mooney-Rivlin LCE material, defined by \eqref{NLC:eq:Wnc:Fn:MR}, when $\textbf{F}=\mathrm{{diag}}\left(\lambda^{-2},\lambda,\lambda\right)$ and $\textbf{G}=\mathrm{diag}\left(a^{-1/3},a^{1/6},a^{1/6}\right)$, with $a=2$ and $\lambda>a^{1/6}\approx 1.1225$; (c) Mooney-Rivlin LCE material with $a=2$, while the shear modulus $\mu$ follows a Gamma distribution with $\rho_{1}=405$, $\rho_{2}=0.01$, and $R_{1}=\mu_{1}/\mu$ is drawn from a Beta distribution with $\xi_{1}=900$, $\xi_{2}=100$; and (d) Mooney-Rivlin LCE material with $\mu=4.05$ and $R_{1}=\mu_{1}/\mu=0.9$, while the shape parameter $a$ follows a Gamma distribution with $\rho_{1}=200$, $\rho_{2}=0.01$. In (a)-(b), when the parameter ratio $\mu_{1}/\mu$ is sufficiently small, the required stress changes from increasing to decreasing, i.e., the material displays inflation instability. For the nematic model, instability is expected at larger deformation than for the underlying hyperelastic model. In (c)-(d), the dashed black lines correspond to the deterministic solutions based only on the mean values of the parameters, whereas the red versions show the arithmetic mean value solutions. When the material parameters are stochastic, the critical load for instability is found in a probabilistic interval where both stable and unstable states are observed with a given probability.}\label{NLC:fig:inflation-MR}
	\end{center}
\end{figure}

Internally pressurized hollow spheres and tubes are relevant in many biological and engineering structures \cite{goriely17}. For rubber spherical and tubular balloons, the first experimental observations of inflation instabilities under internal pressure were reported in \cite{Mallock:1891}. Cylindrical tubes of homogeneous isotropic incompressible hyperelastic material subject to finite symmetric inflation and stretching were theoretically analyzed for the first time in \cite{Rivlin:1949:VI}. Finite radially symmetric inflation of elastic spherical shells was initially investigated in \cite{Green:1950:GS}, then in \cite{Adkins:1952:AR,Shield:1972}. For both elastic tubular and spherical shells, it was shown in \cite{Carroll:1987} that, depending on the material model, the internal pressure may increase monotonically, or increase and then decrease, or increase, decrease, and then increase again. These classical results were extended to elastic materials with stochastic parameters in \cite{Mihai:2021:MA,Mihai:2019a:MDWG,Mihai:2019b:MDWG}. Recent theoretical investigations of inflated nematic cylindrical balloons were presented in \cite{Giudici:2020:GB,Lee:2021:LB}. 

 To compare \emph{inflation instabilities} in nematic and in purely elastic spheres, we consider the hyperelastic Mooney-Rivlin model \cite{Mooney:1940,Rivlin:1948:IV}, given by
\begin{equation}\label{NLC:eq:Wiso:MR}
W(\textbf{A})=\frac{\mu_{1}}{2}\left[\mathrm{tr}\left(\textbf{A}\textbf{A}^\text{T}\right)-3\right]+\frac{\mu_{2}}{2}\left\{\mathrm{tr}\left[\mathrm{Cof}\left(\textbf{A}\textbf{A}^\text{T}\right)\right]-3\right\},
\end{equation}
where $\mu=\mu_{1}+\mu_{2}>0$ is the shear modulus at infinitesimal strain. A Mooney-Rivlin-type neoclassical strain-energy function for the nematic material then takes the form
\begin{equation}\label{NLC:eq:Wnc:Fn:MR}
W^{(nc)}(\textbf{F},\textbf{n})=W(\textbf{G}^{-1}\textbf{F}).
\end{equation}

Taking a spherical coordinates system with coordinates $(R,\Theta,\Phi)$ in the reference configuration, we assume that the sphere is deformed by radially symmetric inflation with deformation gradient $\textbf{F}=\mathrm{{diag}}\left(\lambda^{-2},\lambda,\lambda\right)$, while the natural deformation tensor is $\textbf{G}=\mathrm{diag}\left(a^{-1/3},a^{1/6},a^{1/6}\right)$, and $\lambda>a^{1/6}>1$.  We denote $\mathcal{W}^{(nc)}(\lambda,\textbf{n})=W^{(nc)}(\textbf{F},\textbf{n})$, and further assume that the shell is thin, i.e., $0<\epsilon=(B-A)/A\ll1$, where $A$ and $B$ represent the inner and outer radii of the reference shell, respectively. When the deformation is due to a radial pressure applied uniformly on the inner surface in the present configuration, the corresponding radial Cauchy stress at the inner surface can be approximated as $T^{(nc)}=\epsilon a^{1/2}\lambda^{-2}\mathrm{d}\mathcal{W}^{(nc)}/\mathrm{d}\lambda$. The relation between the Cauchy stress at the inner surface in the nematic and purely elastic shell with the same shear modulus is $T^{(nc)}=T$. Then Figure~\ref{NLC:fig:inflation-MR}(a)-(b) shows that, if the parameter ratio $\mu_{1}/\mu$ is sufficiently small, the required stress changes from increasing to decreasing, i.e., the material displays inflation instability, and for the nematic model, instability is expected at larger deformation than for the hyperelastic model. However, this value will decrease if the model is modified to include an additional elastic energy as in \eqref{NLC:eq:W:Fn}, so that the elasto-nematic ratio is $\eta>0$, while the shear modulus $\mu$ remains the same. For the nematic shell, in Figure~\ref{NLC:fig:inflation-MR}(c)-(d), either the shear modulus $\mu=\mu_{1}+\mu_{2}$ or the shape parameter $a$ is a random variable. In  both cases, the critical load for instability resides in a probabilistic interval where the stable and unstable states compete. To decrease the chance of stable deformation, one must increase the value of $\mu_{1}/\mu$, and only when $\mu_{1}=\mu$ unstable deformation is certain. 

\section{Necking}

\begin{figure}[htbp]
	\begin{center}
		\subfigure[]{\includegraphics[width=0.49\textwidth]{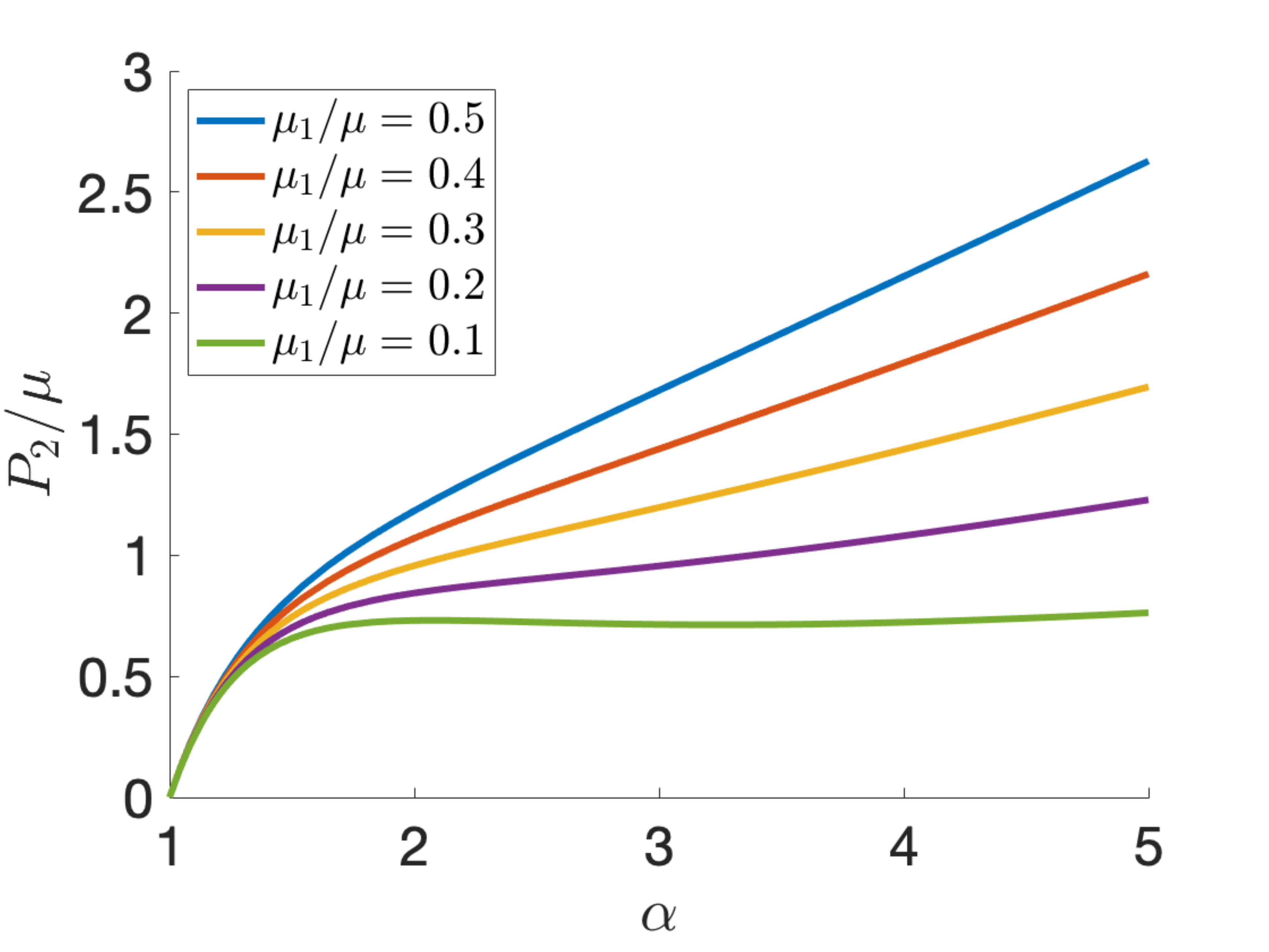}}
		\subfigure[]{\includegraphics[width=0.49\textwidth]{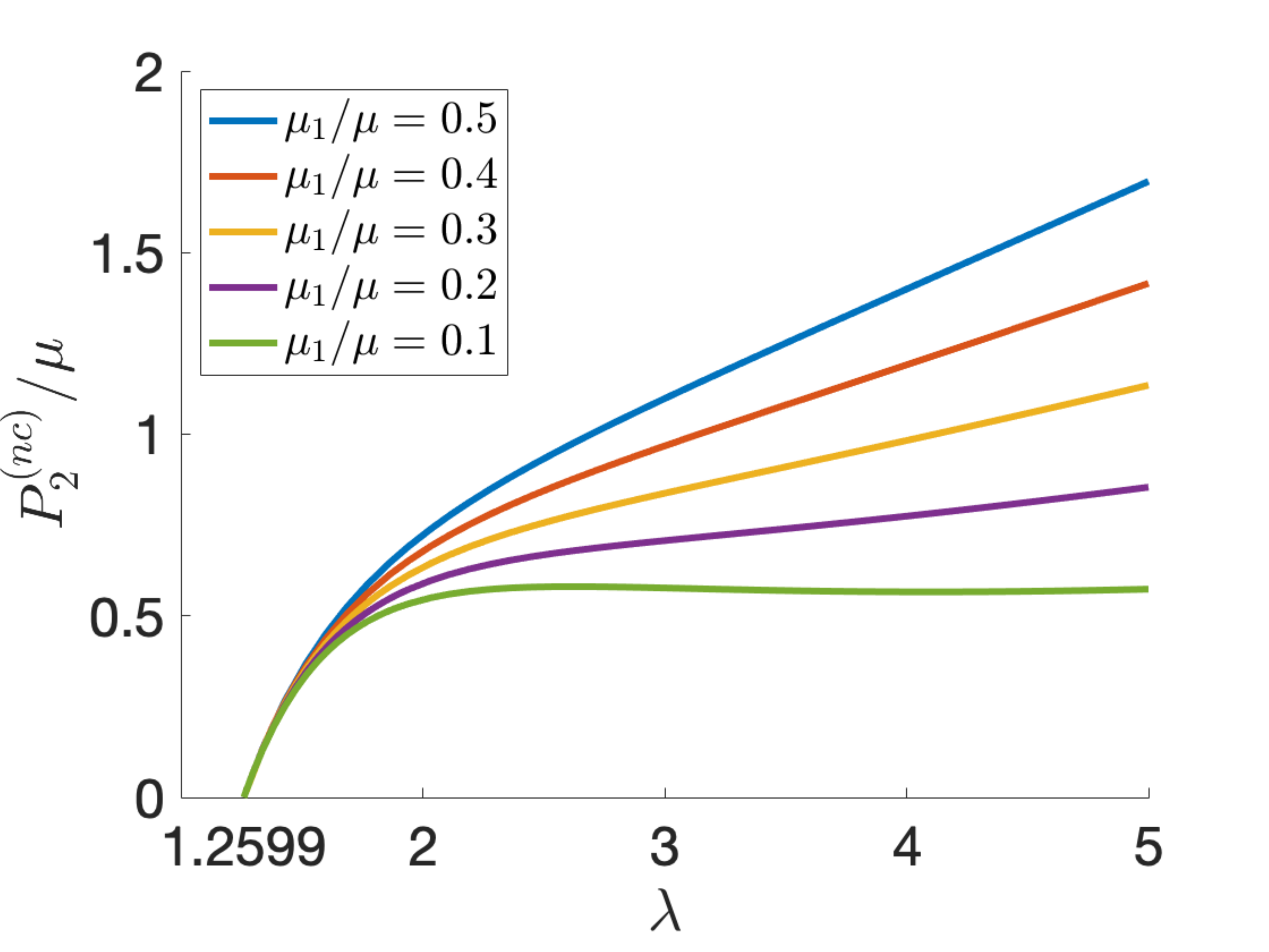}}\\
		\subfigure[]{\includegraphics[width=0.49\textwidth]{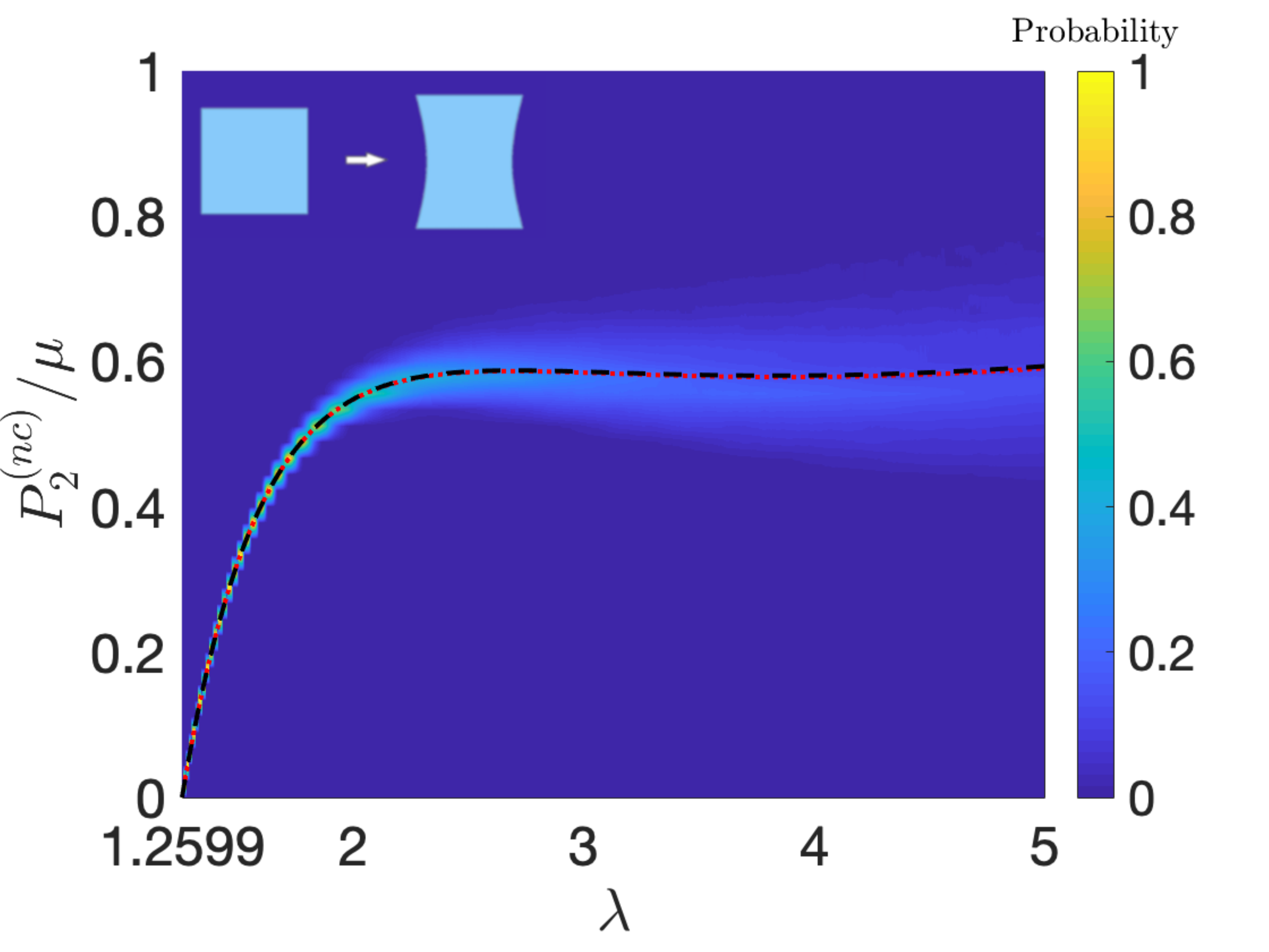}}
		\subfigure[]{\includegraphics[width=0.49\textwidth]{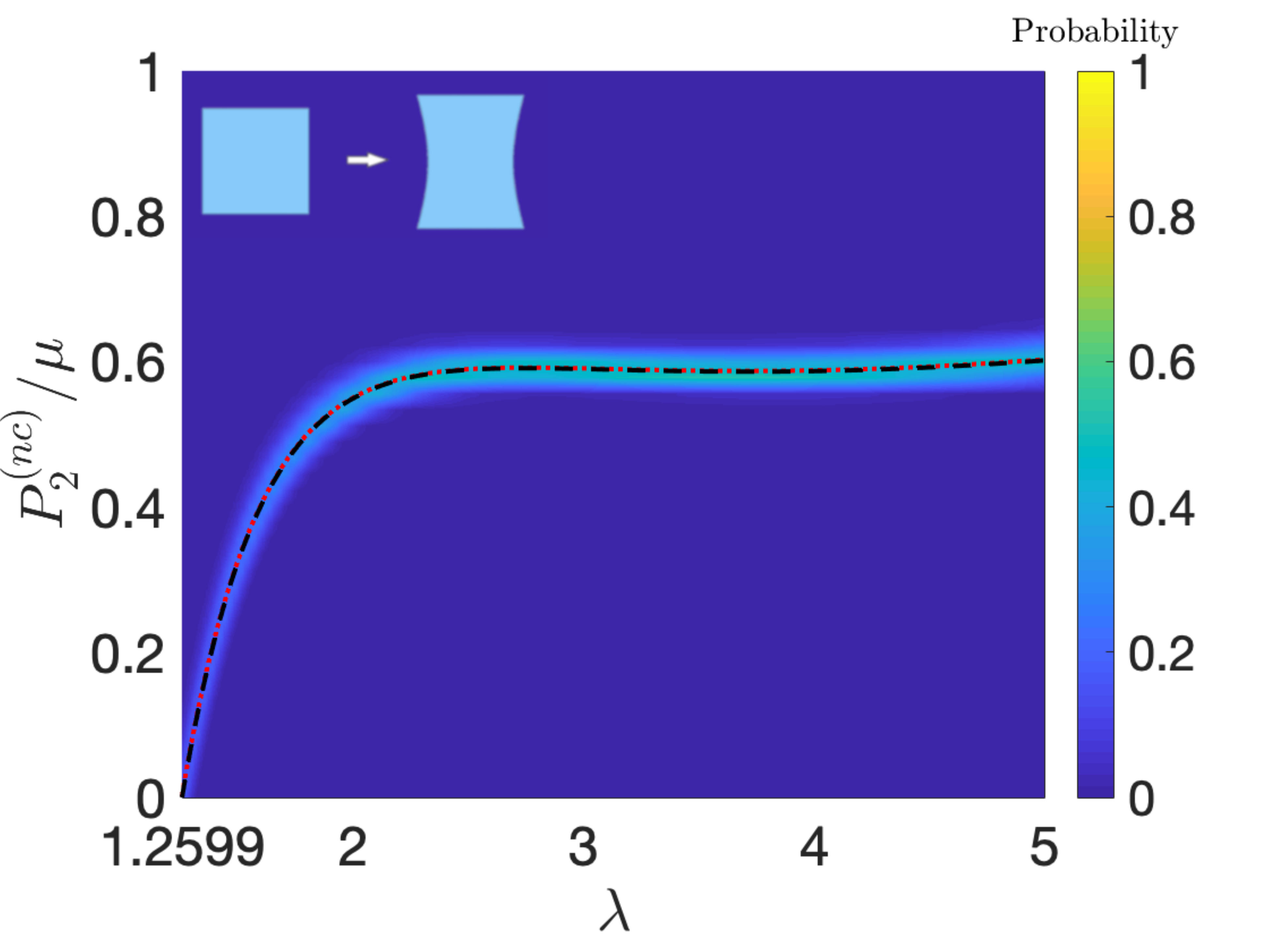}}
		\caption{The effect of changing the value of parameter ratio $\mu_{1}/\mu$ on the normalized tensile first Piola-Kirchhoff stress for: (a) the Gent-Thomas model, defined by \eqref{NLC:eq:Wiso:GT}, when $\textbf{A}=\mathrm{{diag}}\left(\alpha^{-1/2},\alpha,\alpha^{-1/2}\right)$ with $\alpha>1$; (b) the Gent-Thomas-type LCE model, defined by \eqref{NLC:eq:Wnc:Fn:GT}, when $\textbf{F}=\mathrm{{diag}}\left(\lambda^{-1/2},\lambda,\lambda^{-1/2}\right)$ and $\textbf{G}=\mathrm{diag}\left(a^{-1/6},a^{1/3},a^{-1/6}\right)$, with $a=2$ and $\lambda>a^{1/3}\approx 1.2599$; (c) the LCE model when $a=2$, while the shear modulus $\mu$ follows a Gamma distribution with $\rho_{1}=405$, $\rho_{2}=0.01$, and $R_{1}=\mu_{1}/\mu$ is drawn from a Beta distribution with $\xi_{1}=12$, $\xi_{2}=100$; and (b) the LCE model when $\mu=4.05$ and $R_{1}=\mu_{1}/\mu=0.11$, while the shape parameter $a$ follows a Gamma distribution with $\rho_{1}=200$, $\rho_{2}=0.01$. In (c)-(d), the dashed black lines correspond to the deterministic solutions based only on the mean values of the parameters, whereas the red versions show the arithmetic mean value solutions. In all cases, when the parameter ratio is sufficiently small, the required dead load changes from increasing to decreasing, and the material displays necking instability. For the nematic model, necking is expected at larger deformation and lower maximum load than for the underlying hyperelastic model. When the material parameters are stochastic, both stable and unstable states are presented with a given probability.}\label{NLC:fig:necking-GT}
	\end{center}
\end{figure}

Like other rubber-like materials, LCEs may suffer from \emph{necking instability} under stretch \cite{Clarke:1998:CT,Clarke:1998:CTKF}. When homogeneous isotropic incompressible hyperelastic materials are subject to large tension, necking occurs if there is a maximum load, or in other words, if there exists a critical extension ratio, such that the force required to extend the material beyond this critical value changes from increasing to decreasing \cite{Antman:1973,Antman:1977:AC,Audoly:2016:AH,Coleman:1988:CN,Ericksen:1975,Hill:1975:HH,Owen:1987}. The relation between the onset of necking and the maximum load was originally analyzed for ductile materials in \cite{Considere:1885}. For a class of hyperelastic materials where the load-displacement curve does not possess a maximum, in \cite{Sivaloganathan:2011:SS}, it was proved that the homogeneous deformation is the only absolute minimizer of the elastic energy. In particular, incompressible neo-Hookean or Mooney-Rivlin hyperelastic models do not exhibit necking, and this property is inherited by the associated nematic LCE models. For example, in \cite{He:2020:HZHC}, a necking instability observed experimentally under uniaxial tension could not be captured by the neoclassical LCE model based on the neo-Hookean strain-energy function alone. In that case, since necking was initiated during director rotation, a composite model consisting of a purely neoclassical form and an additional elastic form, as presented in Section~\ref{NLC:sec:striping}, should be used to predict the non-zero stress plateau associated with the neck formation.

To further explore necking instability that may occur when the director is parallel to the applied tensile force, and compare that with the same behavior in a purely elastic material, we consider the hyperelastic Gent-Thomas model \cite{Gent:1958:GT} defined by
\begin{equation}\label{NLC:eq:Wiso:GT}
W(\textbf{A})=\frac{\mu_{1}}{2}\left[\mathrm{tr}\left(\textbf{A}\textbf{A}^\text{T}\right)-3\right]+\frac{3\mu_{2}}{2}\ln\frac{\mathrm{tr}\left[\mathrm{Cof}\left(\textbf{A}\textbf{A}^\text{T}\right)\right]}{3},
\end{equation}
where $\mu=\mu_{1}+\mu_{2}>0$ is the shear modulus at small strain. A  Gent-Thomas-type neoclassical strain-energy function for the nematic material then reads
\begin{equation}\label{NLC:eq:Wnc:Fn:GT}
W^{(nc)}(\textbf{F},\textbf{n})=W(\textbf{G}^{-1}\textbf{F}).
\end{equation}

We take $\textbf{F}=\mathrm{{diag}}\left(\lambda^{-1/2},\lambda,\lambda^{-1/2}\right)$, while $\textbf{G}=\mathrm{diag}\left(a^{-1/6},a^{1/3},a^{-1/6}\right)$ and $\lambda>a^{1/3}>1$. In this case, denoting $\mathcal{W}^{(nc)}(\lambda,\textbf{n})=W^{(nc)}(\textbf{F},\textbf{n})$, the uniaxial tensile load is given by the first Piola-Kirchhoff stress $P^{(nc)}_{2}=\mathrm{d}\mathcal{W}^{(nc)}/\mathrm{d}\lambda$, and necking occurs when the ratio $\mu_{1}/\mu$ is sufficiently small. The relation between the first Piola-Kirchhoff tensile stress in the nematic and purely elastic case with the same shear modulus is $P^{(nc)}_{2}=a^{-1/3}P_{2}$. As shown in Figure~\ref{NLC:fig:necking-GT}(a)-(b), for the nematic model, necking is expected at larger deformation and lower maximum dead load than for the hyperelastic model. However, the maximum load will increase if the model is modified to include an additional elastic energy as in \eqref{NLC:eq:W:Fn}, so that the elasto-nematic ratio is $\eta>0$, while the shear modulus $\mu$ remains the same. In Figure~\ref{NLC:fig:necking-GT}(c)-(d), the stochastic tensile load in the deformed LCE is illustrated. 

\section{Conclusion}  

Instabilities in liquid crystalline solids can be of potential interest in a range of applications. Here, we present various situations to understand some new possibilities offered by LCEs. Specifically, we combine for the first time nonlinear elastic instabilities with nematic elastomer theory, within a stochastic setting where the material parameters are probabilistic, and compare the results with those from classical nonlinear elasticity. While shear striping instabilities are specific to LCEs, as they do not occur in rubber, instabilities such as cavitation, shell inflation and necking have been widely studied in the context of purely hyperelastic materials. In recent years, we have reviewed such results and extended them to a stochastic elasticity framework by building directly on the deterministic nonlinear theory. For LCEs, similar interesting phenomena can occur, and we offer a perspective on the general methodology to analyze them. Moreover, each of these instability problems could be further formulated and treated under slightly different hypotheses, where the external forces or the type of nematic material might change to reflect different experimental setups and observations. Our scope is to present some key theoretical ingredients to study this class of problems.

The particular choice of numerical values in our calculations are for illustrative purposes only. To compare the stochastic results with the deterministic ones, we sampled from distributions where the parameters have mean values corresponding to the deterministic system. The stochastic results mostly follow the deterministic ones, but transform a single critical value for instability into a large probability of an instability taking place close to that value. However, both stable and unstable states have a quantifiable chance to be observed with a given probability, and small variations in the input model parameters can have a significant impact on whether an instability occurs or not at a certain load. Moreover, we observed that, as deformation progresses, the solution variance tends to change non-uniformly around the mean value, suggesting that the average value may be less significant from a physical point of view if fluctuations become large. 

The results presented here are universal in the sense that they hold for families of LCE models for which their classic hyperelastic counterparts exhibit a similar instability. To gain significant insight into the macroscopic mechanical properties of these materials, it is imperative that they are analyzed and tested under more complex multiaxial deformations and loads before they can be incorporated into real industrial systems. Therefore, we hope that the general field of instabilities for LCE materials may serve as an inspiration for new devices or for systematic testing. For example, in \cite{He:2020:HZWHC} inflation instabilities were reported for elongated nematic balloons. The inflation of cylindrical elastic balloons is often accompanied by different types of instabilities, such as bulging and necking. These phenomena have been analyzed theoretically for hyperelastic balloons in \cite{Fu:2016:FLF,Fu:2021:FLG,Ye:2020:YLF}, but remain virgin territory for LCEs. The possibility of activation of LCEs may finally fulfill the early promises of the rational mechanics pioneers who first demonstrated the existence of nonlinear elastic instabilities and understood that they could be used to design new devices.

\appendix
\setcounter{equation}{0}
\renewcommand{\theequation}{A.\arabic{equation}}
\section{Stochastic model parameters}

We refer to \cite{Guilleminot:2017:GS,Guilleminot:2020,Soize:2017} for extensive reviews on the information-theoretic approach in stochastic elasticity, and to the various papers cited in the main text for numerous examples and applications of continuum models with stochastic parameters. Here, we adopt the following hypothesis required by the stochastic models for nematic liquid crystal elastomers (LCEs) presented in \cite{Mihai:2020a:MG}: For any given finite deformation, at any point in the material, the shear modulus $\mu>0$, the shape parameter $a>0$, and their inverse, $1/\mu$ and $1/a$, respectively, are second order random variables, i.e., they have finite mean value and finite variance. For the shear modulus $\mu$ (and similarly for $a$), to construct a prior probability law, we note that this assumption is guaranteed by setting the following mathematical expectations:
\begin{eqnarray}\label{eq:Emu1}\begin{cases}
E\left[\mu\right]=\underline{\mu}>0,&\\
E\left[\log\ \mu\right]=\nu,& \mbox{such that $|\nu|<+\infty$}.\label{eq:Emu2}\end{cases}
\end{eqnarray}
The first constraint in \eqref{eq:Emu1} specifies the mean value for the random shear modulus $\mu$, while the second constraint provides a condition from which it follows that $1/\mu$ is a second order random variable. Then, by the maximum entropy principle, the shear modulus $\mu$ with mean value $\underline{\mu}$ and standard deviation $\|\mu\|=\sqrt{\text{Var}[\mu]}$ (defined as the square root of the variance, $\text{Var}[\mu]$) follows a Gamma probability distribution with shape and scale parameters $\rho_{1}>0$ and $\rho_{2}>0$ respectively, such that
\begin{equation}\label{eq:rho12}
\underline{\mu}=\rho_{1}\rho_{2},\qquad
\|\mu\|=\sqrt{\rho_{1}}\rho_{2}.
\end{equation}
The corresponding probability density function takes the form 
\begin{equation}\label{eq:mu:gamma}
g(\mu;\rho_{1},\rho_{2})=\frac{\mu^{\rho_{1}-1}e^{-\mu/\rho_{2}}}{\rho_{2}^{\rho_{1}}\Gamma(\rho_{1})},\qquad\mbox{for}\ \mu>0\ \mbox{and}\ \rho_{1}, \rho_{2}>0,
\end{equation}
where $\Gamma:\mathbb{R}^{*}_{+}\to\mathbb{R}$ is the complete Gamma function
\begin{equation}\label{eq:gamma}
\Gamma(z)=\int_{0}^{+\infty}t^{z-1}e^{-t}\text dt.
\end{equation}
The word `hyperparameters' is often used for $\rho_{1}$ and $\rho_{2}$ to distinguish them from $\mu$ and other material constants. 

When $\mu=\mu_{1}+\mu_{2}$, setting a fixed constant value $b>-\infty$, such that $\mu_{i}>b$, $i=1,2$ (e.g., $b=0$ if $\mu_{1}>0$ and $\mu_{2}>0$, although $b$ is not unique in general), we define the auxiliary random variable 
\begin{equation}\label{eq:R12:b}
R_{1}=\frac{\mu_{1}-b}{\mu-2b},
\end{equation}
such that $0<R_{1}<1$. Then, the random model parameters can be expressed equivalently as follows,
\begin{equation}\label{eq:mu12:b}
\mu_{1}=R_{1}(\mu-2b)+b,\qquad \mu_{2}=\mu-\mu_{1}=(1-R_{1})(\mu-2b)+b.
\end{equation}
It is reasonable to assume 
\begin{eqnarray}\begin{cases}
E\left[\log\ R_{1}\right]=\nu_{1},& \mbox{such that $|\nu_{1}|<+\infty$},\label{eq:ER1}\\
E\left[\log(1-R_{1})\right]=\nu_{2},& \mbox{such that $|\nu_{2}|<+\infty$},\label{eq:ER2}\end{cases}
\end{eqnarray}
in which case, the random variable $R_{1}$ follows a standard Beta distribution, with hyperparameters $\xi_{1}>0$ and $\xi_{2}>0$ satisfying
\begin{equation}\label{eq:xi12}
\underline{R}_{1}=\frac{\xi_{1}}{\xi_{1}+\xi_{2}},\qquad
\text{Var}[R_{1}]=\frac{\xi_{1}\xi_{2}}{\left(\xi_{1}+\xi_{2}\right)^2\left(\xi_{1}+\xi_{2}+1\right)},
\end{equation}
where $\underline{R}_{1}$ is the mean value and $\text{Var}[R_{1}]$ is the variance of $R_{1}$. The associated probability density function is 
\begin{equation}\label{eq:betaR1}
\beta(r;\xi_{1},\xi_{2})=\frac{r^{\xi_{1}-1}(1-r)^{\xi_{2}-1}}{B(\xi_{1},\xi_{2})},\qquad \qquad\mbox{for}\ r\in(0,1)\ \mbox{and}\ \xi_{1}, \xi_{2}>0,
\end{equation}
where $B:\mathbb{R}^{*}_{+}\times\mathbb{R}^{*}_{+}\to\mathbb{R}$ is the Beta function
\begin{equation}\label{eq:beta}
B(x,y)=\int_{0}^{1}t^{x-1}(1-t)^{y-1}dt.
\end{equation}
Then, for the random coefficients given by \eqref{eq:mu12:b}, the corresponding mean values are
\begin{equation}\label{eq:mu12:mean}
\underline{\mu}_{1}=\underline{R}_{1}(\underline{\mu}-2b)+b,
\qquad \underline{\mu}_{2}=\underline{\mu}-\underline{\mu}_{1}=(1-\underline{R}_{1})(\underline{\mu}-2b)+b,
\end{equation}
and the variances and covariance take the form, respectively,
\begin{eqnarray}
&&\text{Var}\left[\mu_{1}\right]
=(\underline{\mu}-2b)^2\text{Var}[R_{1}]+(\underline{R}_{1})^2\text{Var}[\mu]+\text{Var}[\mu]\text{Var}[R_{1}],\\
&&\text{Var}\left[\mu_{2}\right]
=(\underline{\mu}-2b)^2\text{Var}[R_{1}]+(1-\underline{R}_{1})^2\text{Var}[\mu]+\text{Var}[\mu]\text{Var}[R_{1}],\\
&&\text{Cov}[\mu_{1},\mu_{2}]=\frac{1}{2}\left(\text{Var}[\mu]-\text{Var}[\mu_{1}]-\text{Var}[\mu_{2}]\right).
\end{eqnarray}

\setcounter{equation}{0}
\renewcommand{\theequation}{B.\arabic{equation}}
\section{Stress tensors for ideal nematic elastomers}

We briefly recall the relations between the stress tensors of an ideal nematic elastomer and those of the underlying hyperelastic model. These relations were originally obtained in \cite{Mihai:2020b:MG}. For an ideal incompressible nematic elastomer, the neoclassical strain-energy density function takes the general form
\begin{equation}\label{NLC:eq:Wnc}
W^{(nc)}(\textbf{F},\textbf{n})=W(\textbf{A}),
\end{equation}
where the right-hand side represents the strain-energy function of a homogeneous isotropic incompressible hyperelastic material, depending only on the elastic deformation gradient $\textbf{A}$. On the left-hand side, $\textbf{n}$ is a unit vector for the localized direction of uniaxial nematic alignment in the present configuration; $\textbf{F}=\textbf{G}\textbf{A}$ is the deformation gradient tensor with respect to the reference isotropic state, with $\textbf{G}=a^{-1/6}\textbf{I}+\left(a^{1/3}-a^{-1/6}\right)\textbf{n}\otimes\textbf{n}$ the  `spontaneous' (or `natural') deformation tensor and $\textbf{A}$ the (local) elastic deformation tensor; $a>0$ is a temperature-dependent, spatially-independent shape parameter; $\otimes$ denotes the tensor product of two vectors; and $\textbf{I}=\text{diag}(1,1,1)$ is the identity tensor. 

The strain-energy function given by \eqref{NLC:eq:Wnc} takes the equivalent form
\begin{equation}\label{NLC:eq:Wnc:lambdas}
\mathcal{W}^{(nc)}(\lambda_{1},\lambda_{2},\lambda_{3},\textbf{n})=W^{(nc)}(\textbf{F},\textbf{n}),
\end{equation}
where $\{\lambda_{i}^2\}_{i=1,2,3}$ are the eigenvalues of the tensor $\textbf{F}\textbf{F}^{T}$.

For the hyperelastic material described by the strain-energy function $W(\textbf{A})$, the Cauchy stress tensor (representing the internal force per unit of deformed area acting within the deformed solid) is equal to
\begin{equation}\label{NLC:eq:cauchy:iso}
\textbf{T}=\left(\det\textbf{A}\right)^{-1}\frac{\partial W}{\partial\textbf{A}}\textbf{A}^{T}-p\textbf{I},
\end{equation}
where $p$ denotes the Lagrange multiplier for the incompressibility constraint $\det\textbf{A}=1$.

The associated first Piola-Kirchhoff stress tensor (representing the internal force per unit of undeformed area acting within the deformed solid) is then
\begin{equation}\label{NLC:eq:1PK:iso}
\textbf{P}=\textbf{T}\text{Cof}(\textbf{A}),
\end{equation}
where $\text{Cof}(\textbf{A})=\left(\det\textbf{A}\right)\textbf{A}^{-T}$ is the cofactor of $\textbf{A}$. 

\subsection{Free director}

When the nematic director is `free' to rotate relative to the elastic matrix, $\textbf{F}$ and $\textbf{n}$ are independent variables, and the Cauchy stress tensor for the nematic material with the strain-energy function described by \eqref{NLC:eq:Wnc} is calculated as follows,
\begin{equation}\label{NLC:eq:cauchy}
\begin{split}
\textbf{T}^{(nc)}
&=J^{-1}\frac{\partial W^{(nc)}}{\partial\textbf{F}}\textbf{F}^{T}-p^{(nc)}\textbf{I}\\
&=J^{-1}\textbf{G}^{-1}\frac{\partial W}{\partial\textbf{A}}\textbf{A}^{T}\textbf{G}-p^{(nc)}\textbf{I}\\
&=J^{-1}\textbf{G}^{-1}\textbf{T}\textbf{G},
\end{split}
\end{equation}
where $\textbf{T}$ is the Cauchy stress tensor defined by \eqref{NLC:eq:cauchy:iso}, $J=\det\textbf{F}$, and the scalar $p^{(nc)}$ represents the Lagrange multiplier for the internal constraint $J=1$.

The principal components $\left(T^{(nc)}_{1},T^{(nc)}_{2},T^{(nc)}_{3}\right)$ of the Cauchy stress defined by \eqref{NLC:eq:cauchy} are the solutions of the characteristic equation
\begin{equation}
\det\left(\textbf{T}^{(nc)}-\Lambda\textbf{I}\right)=0.
\end{equation}
Since
\begin{equation}
\det\left(\textbf{T}^{(nc)}-\Lambda\textbf{I}\right)=\det\left[\textbf{G}^{-1}\left(J^{-1}\textbf{T}-\Lambda\textbf{I}\right)\textbf{G}\right]=J^{-1}\det\left(\textbf{T}-J\Lambda\textbf{I}\right),
\end{equation}
it follows that the principal Cauchy stresses for the underlying hyperelastic model satisfy
\begin{equation}
\left(T_{1},T_{2},T_{3}\right)=J\left(T^{(nc)}_{1},T^{(nc)}_{2},T^{(nc)}_{3}\right). 
\end{equation}
Therefore, if the Baker-Ericksen inequalities hold for the hyperelastic model, then the greater principal Cauchy stress occurs in the direction of the greater principal elastic stretch for the nematic model. We recall that, for a hyperelastic material, the Baker-Ericksen inequalities state that the greater principal stress occurs in the direction of the greater principal stretch \cite{BakerEricksen:1954,Marzano:1983}.

In the presence of a nematic field, the total Cauchy stress tensor $\textbf{T}^{(nc)}$ given by \eqref{NLC:eq:cauchy} is not symmetric in general. In addition, the following condition is required,
\begin{equation}\label{NLC:eq:orient}
\frac{\partial W^{(nc)}}{\partial\textbf{n}}=\textbf{0},
\end{equation}
or equivalently, by the principle of material objectivity,
\begin{equation}\label{NLC:eq:orient:matobj}
\frac{1}{2}\left(\textbf{T}^{(nc)}-\textbf{T}^{(nc)T}\right)\textbf{n}=\textbf{0},
\end{equation}
where $\left(\textbf{T}^{(nc)}-\textbf{T}^{(nc)^{T}}\right)/2$ represents the skew-symmetric part of the Cauchy stress tensor.

The first Piola-Kirchhoff stress tensor for the nematic material is equal to
\begin{equation}\label{NLC:eq:1PK:nc}
\textbf{P}^{(nc)}=\textbf{T}^{(nc)}\text{Cof}(\textbf{F})=\textbf{G}^{-1}\textbf{T}\textbf{A}^{-T}=\textbf{G}^{-1}\textbf{P},
\end{equation}
where $\textbf{P}$ is the first Piola-Kirchhoff stress given by \eqref{NLC:eq:1PK:iso}.

\subsection{Frozen director}

If the nematic director is `frozen', the Cauchy stress tensor for the nematic material takes the form
\begin{equation}\label{NLC:eq:cauchy:frozen}
\widehat{\textbf{T}}^{(nc)}
=J^{-1}\textbf{G}^{-1}\textbf{T}\textbf{G}-J^{-1}q\left(\textbf{I}-\frac{\textbf{F}\textbf{n}_{0}\otimes\textbf{F}\textbf{n}_{0}}{|\textbf{F}\textbf{n}_{0}|^2}\right)\textbf{n}\otimes\frac{\textbf{F}\textbf{n}_{0}}{|\textbf{F}\textbf{n}_{0}|},
\end{equation}
where $\textbf{T}$ is the Cauchy stress defined by \eqref{NLC:eq:cauchy:iso}, $J=\det\textbf{F}$, $p^{(nc)}$ is the Lagrange multiplier for the volume constraint $J=1$, and  $q$ is the Lagrange multiplier for the constraint
\begin{equation}\label{NLC:eq:n0n}
\textbf{n}=\frac{\textbf{F}\textbf{n}_{0}}{|\textbf{F}\textbf{n}_{0}|}.
\end{equation}

As the Cauchy stress tensor given by \eqref{NLC:eq:cauchy:frozen} is not symmetric in general, the following additional condition must hold,
\begin{equation}\label{NLC:eq:orient:frozen}
\frac{\partial\widehat{W}^{(nc)}}{\partial\textbf{n}}=\textbf{0},
\end{equation}
or equivalently,
\begin{equation}\label{NLC:eq:orient:matobj:frozen}
\frac{1}{2}\left(\widehat{\textbf{T}}^{(nc)}-\widehat{\textbf{T}}^{(nc)T}\right)\textbf{n}=\textbf{0}.
\end{equation}
The corresponding first Piola-Kirchhoff stress tensor for the nematic material is equal to
\begin{equation}\label{NLC:eq:1PK:frozen}
\widehat{\textbf{P}}^{(nc)}=\widehat{\textbf{T}}^{(nc)}\text{Cof}(\textbf{F}).
\end{equation}

\setcounter{equation}{0}
\renewcommand{\theequation}{C.\arabic{equation}}
\section{Cavitation of a nematic sphere}

The static and dynamic cavitation of homogeneous and radially inhomogeneous isotropic incompressible hyperelastic spheres with stochastic material parameters was analyzed in \cite{Mihai:2019c:MDWG,Mihai:2020:MWG} where up-to-date reviews of the relevant literature are presented. For an inflated elastic sphere, the radially symmetric deformation takes the form
\begin{equation}\label{eq:sphere:deform}
r=g(R),\qquad \theta=\Theta,\qquad \phi=\Phi,
\end{equation}
where $(R,\Theta,\Phi)$ and $(r,\theta,\phi)$ are the spherical polar coordinates in the reference and current configuration, respectively, such that $0\leq R\leq B$, and $g(R)\geq 0$ is to be determined. The corresponding deformation gradient is equal to $\textbf{A}=\mathrm{diag}\left(\alpha_{1},\alpha_{2},\alpha_{3}\right)$, with
\begin{equation}\label{eq:sphere:lambdas:g}
\alpha_{1}=\frac{\mathrm{d}g}{\mathrm{d}R}=\alpha^{-2},\qquad \alpha_{2}=\alpha_{3}=\frac{g(R)}{R}=\frac{r}{R}=\alpha,
\end{equation}
where $\alpha_{1}$ and $\alpha_{2}=\alpha_{3}$ are the radial and hoop principal stretches, respectively, and $\mathrm{d}g/\mathrm{d}R$ denotes the derivative of $g$ with respect to $R$. By \eqref{eq:sphere:lambdas:g},
\begin{equation}\label{eq:sphere:dg}
g^2\frac{\mathrm{d}g}{\mathrm{d}R}=R^2,
\end{equation}
hence,
\begin{equation}\label{eq:sphere:g}
g(R)=\left(R^3+c^3\right)^{1/3},
\end{equation}
where $c\geq 0$ is a constant to be calculated. If $c>0$, then $g(R)\to c>0$ as $R\to 0_{+}$, and a spherical cavity of radius $c$ forms at the center of the sphere, from zero initial radius, otherwise the sphere remains undeformed.

In particular, for a static sphere of neo-Hookean material, if the surface of the cavity is traction-free, the radial component of the Cauchy stress is equal to (see \cite{Mihai:2020:MWG} for a detailed derivation using the same notation)
\begin{equation}\label{eq:sphere:T:NH}
T_{rr}(b)=\frac{2}{3}\int_{x^3+1}^{\infty}\mu\frac{1+u}{u^{7/3}}\mathrm{d}u,
\end{equation}
where $x=c/B$. After evaluating the integral in \eqref{eq:sphere:T:NH}, the required uniform dead-load traction at the outer surface, $R=B$, in the reference configuration, takes the form
\begin{equation}\label{eq:sphere:P:NH}
P=\left(x^{3}+1\right)^{2/3}T_{rr}(b)=2\mu\left[\left(x^{3}+1\right)^{1/3}+\frac{1}{4\left(x^{3}+1\right)^{2/3}}\right],
\end{equation}
and increases as $x$ increases. The critical dead load for the onset of cavitation is then
\begin{equation}\label{eq:sphere:P0:NH}
P_{0}=\lim_{x\to0_{+}}P=\frac{5\mu}{2}.
\end{equation}
To analyze the stability of this cavitation, we study the behavior of the cavity opening, with radius $c$ as a function of $P$, in a neighborhood of $P_{0}$. After differentiating the function given by \eqref{eq:sphere:P:NH}, with respect to the dimensionless cavity radius $x=c/B$, we have
\begin{equation}\label{eq:sphere:dP:NH}
\frac{\mathrm{d}P}{\mathrm{d}x}=2\mu x^2\left[\frac{1}{\left(x^{3}+1\right)^{2/3}}-\frac{1}{2\left(x^{3}+1\right)^{5/3}}\right]>0,
\end{equation}
i.e., the cavitation is stable, regardless of the material parameter $\mu>0$.

For a nematic sphere of neoclassical material with the strain-energy function given by \eqref{NLC:eq:Wnc} derived from the neo-Hookean hyperelastic model, if $\textbf{F}=\mathrm{{diag}}\left(\lambda^{-2},\lambda,\lambda\right)$ and $\textbf{G}=\mathrm{diag}\left(a^{-1/3},a^{1/6},a^{1/6}\right)$, with $\lambda>a^{1/6}>1$, then the Cauchy stress is equal to that of the neo-Hookean sphere. Hence, $T^{(nc)}_{rr}=T_{rr}$, and the first Piola-Kirchhoff stress representing the critical dead load for cavitation in the nematic sphere is equal to $P^{(nc)}_{0}=a^{1/3}P_{0}=5a^{1/3}\mu/2$.

\setcounter{equation}{0}
\renewcommand{\theequation}{D.\arabic{equation}}
\section{Inflation of a nematic spherical shell}

Static and dynamic inflation instabilities of homogeneous and radially inhomogeneous isotropic incompressible hyperelastic spheres with stochastic material parameters were analyzed in \cite{Mihai:2021:MA,Mihai:2019a:MDWG,Mihai:2019b:MDWG}. For a thin hyperelastic spherical shell, such that $0<\epsilon=(B-A)/A\ll 1$, where $A$ and $B$ represent the inner and outer radii of the reference shell, respectively, if the external pressure is equal to zero, then the internal pressure can be approximated as follows,
\begin{equation}\label{eq:shell:T}
T=\frac{\epsilon}{\alpha^2}\frac{\text{d}\mathcal{W}}{\text{d}\alpha},
\end{equation}
where the deformation gradient for radially symmetric inflation is equal to $\textbf{A}=\mathrm{{diag}}\left(\alpha^{-2},\alpha,\alpha\right)$, with $\alpha=r/R>1$, and $\mathcal{W}(\alpha)=W(\textbf{A})$. The critical value of $\alpha$ where a limit-point instability occurs is obtained by solving for $\alpha>1$ the following equation,
\begin{equation}\label{eq:shell:lps}
\frac{\text{d}T}{\text{d}\alpha}=0,
\end{equation}
where $T$ is the radial component of the Cauchy stress given by \eqref{eq:shell:T}.

For a nematic sphere of neoclassical material with the strain-energy function given by \eqref{NLC:eq:Wnc} derived from the Mooney-Rivlin hyperelastic model, when $\textbf{F}=\mathrm{{diag}}\left(\lambda^{-2},\lambda,\lambda\right)$ and $\textbf{G}=\mathrm{diag}\left(a^{-1/3},a^{1/6},a^{1/6}\right)$, with $\lambda>a^{1/6}>1$, the Cauchy stress is equal to that of the hyperelastic sphere. If the shell is thin, assuming zero external pressure, the internal pressure can be approximated as
\begin{equation}\label{eq:shell:Tnc}
T^{(nc)}=T=\frac{\epsilon}{\alpha^2}\frac{\text{d}\mathcal{W}}{\text{d}\alpha}=\frac{\epsilon a^{1/2}}{\lambda^{2}}\frac{\mathrm{d}\mathcal{W}^{(nc)}}{\mathrm{d}\lambda}.
\end{equation}
Then, the critical value of $\lambda$ where a limit-point instability occurs is found by solving for $\lambda>a^{1/3}$ the equation
\begin{equation}\label{eq:shell:lps:nc}
\frac{\text{d}T^{(nc)}}{\text{d}\lambda}=0,
\end{equation}
with $T^{(nc)}$ given by \eqref{eq:shell:Tnc}.

\paragraph{Acknowledgement.}  We gratefully acknowledge the support by the Engineering and Physical Sciences Research Council of Great Britain under research grants EP/R020205/1 to Alain Goriely and EP/S028870/1 to L. Angela Mihai.


\end{document}